\RequirePackage{lineno}
\documentclass[aps,titlepage,12pt,superscriptaddress]{revtex4}

\usepackage{multirow}
\usepackage{lettrine}
\usepackage{epstopdf}
\usepackage{epsfig}
\usepackage{float}
\usepackage{graphicx}
\usepackage{times}
\usepackage{amsmath}
\usepackage{color}

\begin{document}

\title{Optimization of institutional incentives for cooperation in structured populations}

\author{Shengxian Wang}
\affiliation{School of Mathematical Sciences, University of Electronic Science and Technology of China, Chengdu 611731, China}
\affiliation{Faculty of Science and Engineering, University of Groningen, Groningen 9747 AG, The Netherlands}

\author{Xiaojie Chen}
\email{xiaojiechen@uestc.edu.cn}
\affiliation{School of Mathematical Sciences, University of
Electronic Science and Technology of China, Chengdu 611731, China}

\author{Zhilong Xiao}
\affiliation{School of Mathematical Sciences, University of Electronic Science and Technology of China, Chengdu 611731, China}
\affiliation{School of Computer Science and Engineering, Sun Yat-sen University,
Guangzhou 510006, China}

\author{Attila Szolnoki}
\affiliation{Institute of Technical Physics and Materials Science, Centre for Energy Research, P.O. Box 49, Budapest H-1525, Hungary}

\author{V\'{\i}tor V. Vasconcelos}
\affiliation{Computational Science Lab, Informatics Institute, University of Amsterdam, 1098XH Amsterdam, The Netherlands}
\affiliation{Institute for Advanced Study, University of Amsterdam, 1012 GC, Amsterdam, The Netherlands}

\begin{abstract}\noindent
\\
The application of incentives, such as reward and punishment, is a frequently applied way for promoting cooperation among interacting individuals in structured populations. However, how to properly use the incentives is still a challenging problem for incentive-providing institutions. In particular, since the implementation of incentive is costly, to explore the optimal incentive protocol, which ensures the desired collective goal at a minimal cost, is worthy of study. In this work, we consider the positive and negative incentives respectively for a structured population of individuals whose conflicting interactions are characterized by a prisoner's dilemma game. We establish an index function for quantifying the cumulative cost during the process of incentive implementation, and theoretically derive the optimal positive and negative incentive protocols for cooperation on regular networks. We find that both types of optimal incentive protocols are identical and time-invariant. Moreover, we compare the optimal rewarding and punishing schemes concerning implementation cost and provide a rigorous basis for the usage of incentives in the game-theoretical framework. We further perform computer simulations to support our theoretical results and explore their robustness for different types of population structures, including regular, random, small-world, and scale-free networks.
\end{abstract}

\maketitle

%\noindent
%\linenumbers

\leftline{\textbf{1. Introduction}}

Cooperation is of vital importance in the contemporary era~\cite{Hauert_2010}. However, its evolution and emergence conflict with the immediate self-interest of interacting individuals~\cite{Perc17PR}. Evolutionary game theory provides a common mathematical framework to depict how agents interact with each other in a dynamical process and to predict how cooperative action evolves from a population level \cite{Hofbauer_98}. As a representative paradigm, the prisoner's dilemma game has received considerable attention for studying the problem of cooperation in a population of interacting individuals \cite{Nowak_92}.

When individuals play the evolutionary prisoner's dilemma game in a large well-mixed population, in which all are equally likely to interact, cooperation cannot emerge. However, the real-world population structures are not well-mixed but relatively complicated, and the interactions among individuals are limited to a set of neighbors in a structured population~\cite{Erd_59}. Inspired by the fast progress of network science, several interaction topologies have been tested, including small-world \cite{Watts_98} and scale-free networks \cite{Barab_99}. Features of these networks can influence the evolutionary dynamics of cooperation significantly \cite{Santos05PRL,Szab_07, Tarnita09JTB,Allen_B17,Li20NC,Su22NHB,Su22PNAS}. Beside network topology, another crucial determinant of the evolution of cooperation is the strategy update rule, which determines the microscopic update procedure \cite{ohtsuki_h06,ohtsuki_1,Nowak_t10,Zhou21NC}. For instance, cooperation can be favored under the so-called death-birth strategy update rule if the benefit-to-cost ratio in the prisoner's dilemma game exceeds the average degree of the specific interaction network \cite{ohtsuki_h06, ohtsuki_1, Nowak_t10}. In contrast, it cannot emerge under the alternative birth-death strategy update rule \cite{ohtsuki_h06}.

Overall, situations abound when cooperation cannot emerge in structured populations when no additional regulation mechanisms or moral nudges are incorporated \cite{ohtsuki_h06, Capraro21Interface, Capraro22arXiv}. In these unfavorable environments, prosocial incentives can be used to sustain cooperation among unrelated and competing agents \cite{David_09,Henrich_06,Gurerk_06,Sigmund_10,Mann_R17,Dreber_A08,Riehl_J18,Han15Interface,Vasconcelos_NS22}. Specifically, cooperators can be rewarded for their positive acts or defectors are punished for their sweepingly negative impact. At the individual level, cooperation can be favored whenever the amount of incentive exceeds the payoff difference between cooperating and defecting, no matter whether the incentive is positive or negative. However, in a given unfavorable environment for the emergence of cooperation, it is still unclear how intensive positive or negative incentive is needed to drive the population toward the desired direction. Furthermore, applying incentives is always costly \cite{Sasaki_12, Vasconcelos_13, Chen_J15, Wang_X19,Duong21PRSA}. Previous related works assume that, when used, the incentive amount is fixed at a certain value~\cite{Sasaki_12,Chen_J15}. In those circumstances, the obtained incentive protocol is not necessarily the one with the minimal cumulative cost~\cite{Wang_X19}. Therefore, finding the optimal time-varying incentive protocol is vital to ensure effective interventions that drive populations towards a productive and cooperative state at a minimal execution cost.

Our work addresses how much incentive is needed for cooperation to emerge and explores the optimal incentive protocols in a game-theoretical framework, where time-varying institutional positive or negative incentives are provided for a structured population of individuals playing the prisoner's dilemma game. We establish an index function for quantifying the executing cumulative cost.
We systematically survey all relevant strategy update rules, including death-birth (DB), birth-death (BD), imitation (IM), and pairwise-comparison (PC) updating \cite{ohtsuki_h06, ohtsuki_1, Nowak_t10, Szabo_98}. Using optimal control theory, we obtain the dynamical incentive protocol under each strategy update rule leading to the minimal cumulative cost for the emergence of cooperation. Interestingly, we find that the optimal negative and positive incentive protocols are identical and time-invariant for each given strategy update rule. However, applying punishment can induce a lower cumulative cost than the usage of reward if the initial cooperation level is larger than the difference between the full cooperation state and the desired cooperation state. Otherwise, applying reward requires a lower cost. Beside analytical calculations, we perform computer simulations confirming that our results are valid in a broad class of population structures.

\leftline{\textbf{2. Results}}

\begin{figure}[ht]
\includegraphics[width=1\textwidth]{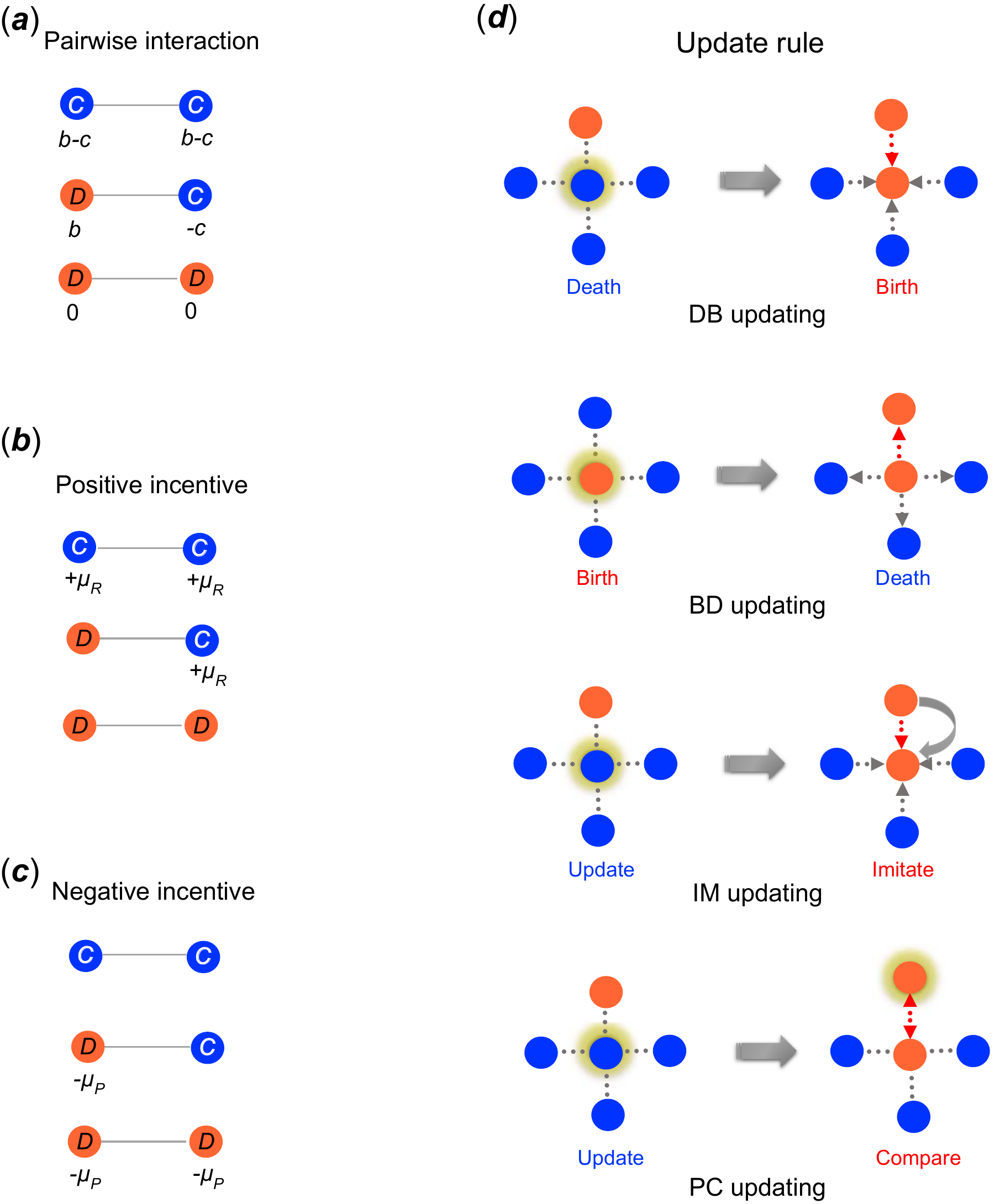}
\caption{\textbf{Evolutionary prisoner's dilemma game on a graph with institutional reward or punishment.} Panel a shows the pairwise interaction between two connected neighbors in a network. Panel b (c) shows how incentives are implemented for the two connected agents who played the game when positive (negative) incentives from the incentive-providing institution are applied. Panel d shows the illustration of the four strategy update rules, depicting how agents update their strategies after obtaining payoffs from the pairwise interactions with neighbors and the incentive-providing institution.
}\label{fig1}
\end{figure}

\noindent
We start with a structured population of $N$ individuals who interact in a regular network of degree $k>2$. In this graph, vertices represent interacting agents and the edges determine who interacts with whom. Each individual $i$ plays the prisoner's dilemma game with its neighbors. An agent can choose either to be a cooperator ($C$), which confers a benefit $b$ to its opponent at a cost $c$ to itself, or to be a defector ($D$), which is costless and does not distribute any benefits. After playing the game with one neighbor, if institutional positive incentives are in place, the agent is rewarded an amount $\mu_R$ of incentive for choosing $C$. When negative institutional incentives are implemented, it is fined by an amount $\mu_P$ for choosing $D$. The agent collects an accumulated payoff $\pi_i$ by interacting with all neighbors. We set the fitness, $f_i$, of individual $i$, chiefly the reproductive rate, as $1-\omega+\omega \pi_i$, where $\omega$ ($0\leq\omega \leq1$) measures the strength of selection~\cite{Nowak_MA04}. In this work, we concentrate on the effects of weak selection, meaning that $0<\omega \ll 1$. Four different fitness-dependent strategy update rules are considered separately, and the details are given in the Model and Methods section. In addition, in order to help readers intuitively understand the evolutionary process in networked prisoner's dilemma game with institutional reward or punishment, we present an illustration figure as shown in figure~\ref{fig1}.

\vbox{}

\noindent\textbf{2.1. Theoretical predictions for optimal incentive protocols}

In this subsection, we respectively explore the incentive protocols under four alternative strategy update rules, including DB, BD, IM, and PC updating. The details of these calculations are described in electronic supplementary material, and here we only summarize the pair approximation approach for our dynamical system with positive or negative incentive in the weak selection limit.

Using the pair approximation approach (see section~1 in electronic supplementary material), we get the dynamical equation of the fraction of cooperators under DB rule with positive or negative incentive as
\begin{eqnarray}
\frac{dp_{C}}{dt}=F_{\textrm{DB}}(p_{C}, \mu_{v}, t)=\frac{\omega(k-2)}{k-1}[b+k(\mu_{v}-c)]p_{C}(1-p_{C})+o(\omega^{2})\,,\label{eq1}
\label{DB}
\end{eqnarray}
where $p_{C}$ is the fraction of cooperators in the whole population, and $\mu_{v}$ the amount of positive incentive that one cooperator receives from an incentive-providing institution or the amount of negative incentive imposed on a defector by a central institution (note that such incentive is applied for every interaction as explained in the Model and Methods section). Eq.~(\ref{DB}) has two equilibria, one at $p_{C}=0$ and the other at $p_{C}=1$. If $\mu_{v}> c-\frac{b}{k}$, the first is unstable and the second stable, indicating that cooperators prevail over defectors (further details are presented in section~1 of electronic supplementary material). Notably, in the absence of incentives, i.e., $\mu_v=0$, we get back the previously identified $b/c>k$ condition for the evolution of cooperation \cite{ohtsuki_h06, ohtsuki_1}.

Using the Hamilton-Jacobi-Bellman (HJB) equation \cite{Evans_05, Geering_07, Lenhart_05}, in the condition of $\mu_{v}> c-\frac{b}{k}>0$, we obtain analytically the optimal protocol $\mu_{v}^{\ast}=\frac{2(ck-b)}{k}$ both for reward and punishment (see section~1 in electronic supplementary material for details). The optimal rewarding and optimal punishing protocols are both time-invariant and the optimal incentive levels for punishment and reward are identical, namely, $\mu^{\ast}_v=\mu^{\ast}_R=\mu^{\ast}_P$. Accordingly, with the optimal rewarding or punishing protocol, the dynamical system described by Eq.~(\ref{DB}) can be solved and its solution is $p_{C}=\frac{1}{1+\frac{1-p_{0}}{p_{0}} e^{-\beta_{\textrm{DB}}t}}$, where $\beta_{\textrm{DB}}=\frac{\omega(k-2)(ck-b)}{k-1}$ and $p_{0}=p_C(0)>0$ denotes the initial fraction of cooperators in the population.

The cumulative cost produced by the rewarding protocol $\mu_{R}^{*}$ for the dynamical system to reach the expected terminal state $p_{C}(t_{f})$ from the initial state $p_{0}$ is
\begin{eqnarray}
J_{R}^{\ast}=\int^{t_{f}}_{0}\frac{(kNp_{C}\mu_{R}^{\ast})^{2}}{2}dt=\frac{(kN\mu^{\ast}_{R})^{2}}{2\beta_{\textrm{DB}}}[p_{0}+\delta-1+\ln(\frac{1-p_0}{\delta})],
\end{eqnarray}
and, similarly, the cumulative cost produced by the punishing protocol $\mu_{P}^{*}$ becomes
\begin{eqnarray}
J_{P}^{\ast}=\int^{t_{f}}_{0}\frac{(kNp_{D}\mu_{P}^{\ast})^{2}}{2}dt=\frac{(kN\mu_{P}^{\ast})^{2}}{2\beta_{\textrm{DB}}}[p_{0}+\delta-1+\ln(\frac{1-\delta}{p_0})].
\end{eqnarray}
We further find that $J_{R}^{\ast}>J_{P}^{\ast}$ if $p_0>\delta$ and $J_{R}^{\ast}<J_{P}^{\ast}$ if $p_0<\delta$ (see section~1 in electronic supplementary material for details).

For BD updating, the dynamical equation is
 \begin{eqnarray}
\frac{dp_{C}}{dt}=F_{\textrm{BD}}(p_{C}, \mu_{v}, t)=\frac{\omega k(k-2)}{k-1}(\mu_{v}-c)p_{C}(1-p_{C})+o(\omega^{2}).
\label{BDpair}
\end{eqnarray}
We prove that when $\mu_{v}>c$, the system can reach the stable full cooperation state. Naturally, in the absence of incentives we get back the results of Ohtsuki {\it et al.} \cite{ohtsuki_h06, ohtsuki_1}. By solving the HJB equation, the optimal incentive protocol is $\mu_{v}^{\ast}=2c$ both for reward and punishment. The solution of Eq.~(\ref{BDpair}) is $p_{C} =\frac{1}{1+\frac{1-p_{0}}{p_{0}} e^{-\beta_{\textrm{BD}}t}}$, where $\beta_{\textrm{BD}}=\frac{\omega k(k-2)c}{k-1}$ (see section~2 in electronic supplementary material).  Consequently, the cumulative cost to reach the expected terminal state in case of the optimal rewarding protocol is \begin{eqnarray}
J_{R}^{\ast}=\frac{(kN\mu^{\ast}_{R})^{2}}{2\beta_{\textrm{BD}}}[p_{0}+\delta-1+\ln(\frac{1-p_0}{\delta})],
\end{eqnarray}
while for the optimal punishing protocol it is
\begin{eqnarray}
J_{P}^{\ast}=\frac{(kN\mu_{P}^{\ast})^{2}}{2\beta_{\textrm{BD}}}[p_{0}+\delta-1+\ln(\frac{1-\delta}{p_0})].
\end{eqnarray}

In case of IM updating, the dynamical equation becomes
 \begin{eqnarray}
\frac{dp_{C}}{dt}=F_{\textrm{IM}}(p_{C}, \mu_{v}, t)=\frac{\omega k^{2}(k-2)}{(k+1)^{2}(k-1)}[b+(\mu_{v}-c)(k+2)]p_{C}(1-p_{C})+o(\omega^{2})\,,
\label{IMpair}
\end{eqnarray}
which indicates that the system evolves to the full cooperation state if $\mu_{v}>c-\frac{b}{k+2}$. The solution of HJB for the optimal incentive protocol gives $\mu_{v}^{\ast}=\frac{2[c(k+2)-b]}{k+2}$ both for reward and punishment (see section~3 in electronic supplementary material). The solution of Eq.~(\ref{IMpair}) is $p_{C}=\frac{1}{1+\frac{1-p_{0}}{p_{0}} e^{-\beta_{\textrm{IM}}t}}$, where $\beta_{\textrm{IM}}=\frac{\omega k^{2}(k-2)[c(k+2)-b]}{(k+1)^{2}(k-1)}$. The cumulative cost requires for the optimal rewarding protocol is
\begin{eqnarray}
J_{R}^{\ast}=\frac{(kN\mu^{\ast}_{R})^{2}}{2\beta_{\textrm{IM}}}[p_{0}+\delta-1+\ln(\frac{1-p_0}{\delta})],
\end{eqnarray}
while for the optimal punishing protocol it becomes
\begin{eqnarray}
J_{P}^{\ast}=\frac{(kN\mu_{P}^{\ast})^{2}}{2\beta_{\textrm{IM}}}[p_{0}+\delta-1+\ln(\frac{1-\delta}{p_0})].
\end{eqnarray}

When PC updating is applied, the dynamical equation is given by
  \begin{eqnarray}
\frac{dp_{C}}{dt}=F_{\textrm{PC}}(p_{C}, \mu_{v}, t)=\frac{\omega k(k-2)}{2(k-1)}(\mu_{v}-c)p_{C}(1-p_{C})+o(\omega^{2}),
\label{PCpair}
\end{eqnarray}
from which we find that when $\mu_{v}> c$ is satisfied, the stable full cooperation state can be reached (see section~4 in electronic supplementary material). For the optimal incentive protocol, we get $\mu_{v}^{\ast}=\mu_{R}^{\ast}=\mu_{P}^{\ast}=2c$. The solution of Eq.~(\ref{PCpair}) is $p_{C}=\frac{1}{1+\frac{1-p_{0}}{p_{0}} e^{-\beta_{\textrm{PC}}t}}$, where $\beta_{\textrm{PC}}=\frac{\omega k(k-2)c}{2(k-1)}$. Consequently, the cumulative cost for the optimal rewarding protocol is \begin{eqnarray}
J_{R}^{\ast}=\frac{(kN\mu^{\ast}_{R})^{2}}{2\beta_{\textrm{PC}}}[p_{0}+\delta-1+\ln(\frac{1-p_0}{\delta})],
\end{eqnarray}
while it is
\begin{eqnarray}
J_{P}^{\ast}=\frac{(kN\mu_{P}^{\ast})^{2}}{2\beta_{\textrm{PC}}}[p_{0}+\delta-1+\ln(\frac{1-\delta}{p_0})],
\end{eqnarray}
for the optimal punishing protocol.

\begin{figure}[ht]
\includegraphics[width=1\textwidth]{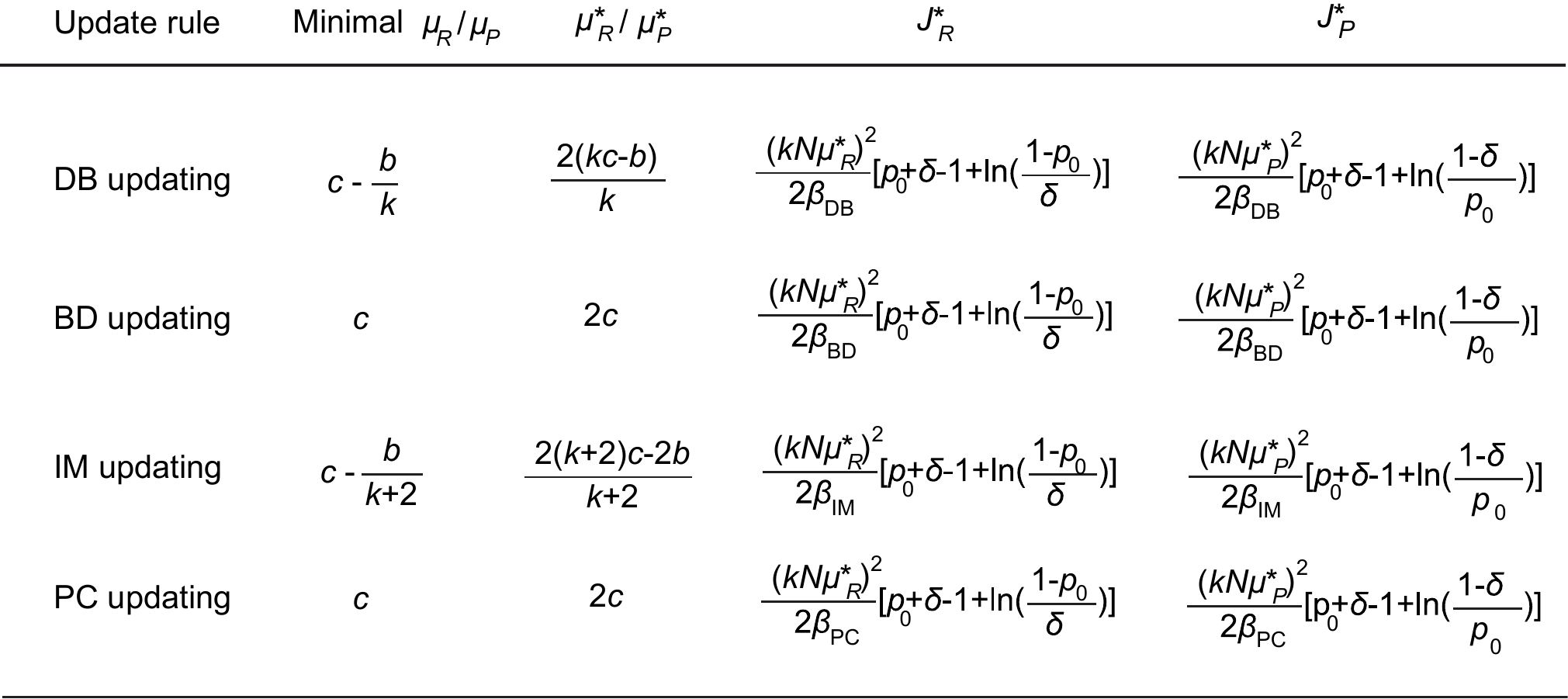}
\caption{\textbf{Optimization of institutional incentives for cooperation for different strategy update rules.} Here, the minimal $\mu_R$ ($\mu_P$) means the minimal amount of positive (negative) incentive needed for the evolution of cooperation for different strategy update rules. $\mu_{R}^{*}$ ($\mu_{P}^{*}$) represents the optimal positive (negative) incentive protocol for different strategy update rules. $J_R^*$ ($J_P^*$) means the cumulative cost produced by the optimal rewarding (punishing) protocol $\mu_{R}^{*}$ ($\mu_{P}^{*}$) for the dynamical system to reach the expected terminal state $1-\delta$ from the initial state $p_{0}$. In addition, $N$ denotes the population size and $k$ the degree of the regular network. $b$ represents the benefit of cooperation and $c$ the cost of cooperation. The parameters $\beta_{\textrm{DB}}=\frac{\omega(k-2)(ck-b)}{k-1}$ under DB updating, $\beta_{\textrm{BD}}=\frac{\omega k(k-2)c}{k-1}$ under BD updating, $\beta_{\textrm{IM}}=\frac{\omega k^{2}(k-2)[c(k+2)-b]}{(k+1)^{2}(k-1)}$ under IM updating, and $\beta_{\textrm{PC}}=\frac{\omega k(k-2)c}{2(k-1)}$ under PC updating.
}\label{fig2}
\end{figure}

Note that for each update rule, the governing equation in the weak selection limit always has two equilibrium points, which are the full defection and full cooperation states, respectively. Accordingly, we can obtain the minimal amounts of incentive needed for the evolution of cooperation, as summarized in figure~\ref{fig2}. Independently of the applied update rule, the optimal incentive protocols are time-invariant and equal both for reward and punishment. Hence, $\mu_{v}^{\ast}=\mu_{R}^{\ast}=\mu_{P}^{\ast}$ for each update rule. The optimal protocols are summarized in figure~\ref{fig2}. We further present the cumulative cost for the optimal reward and punishment protocols for each update rule in figure~\ref{fig2}.

\noindent\textbf{2.2. Numerical calculations and computer simulations for optimal incentive protocols}

\begin{figure}[ht]
\includegraphics[width=0.85\textwidth]{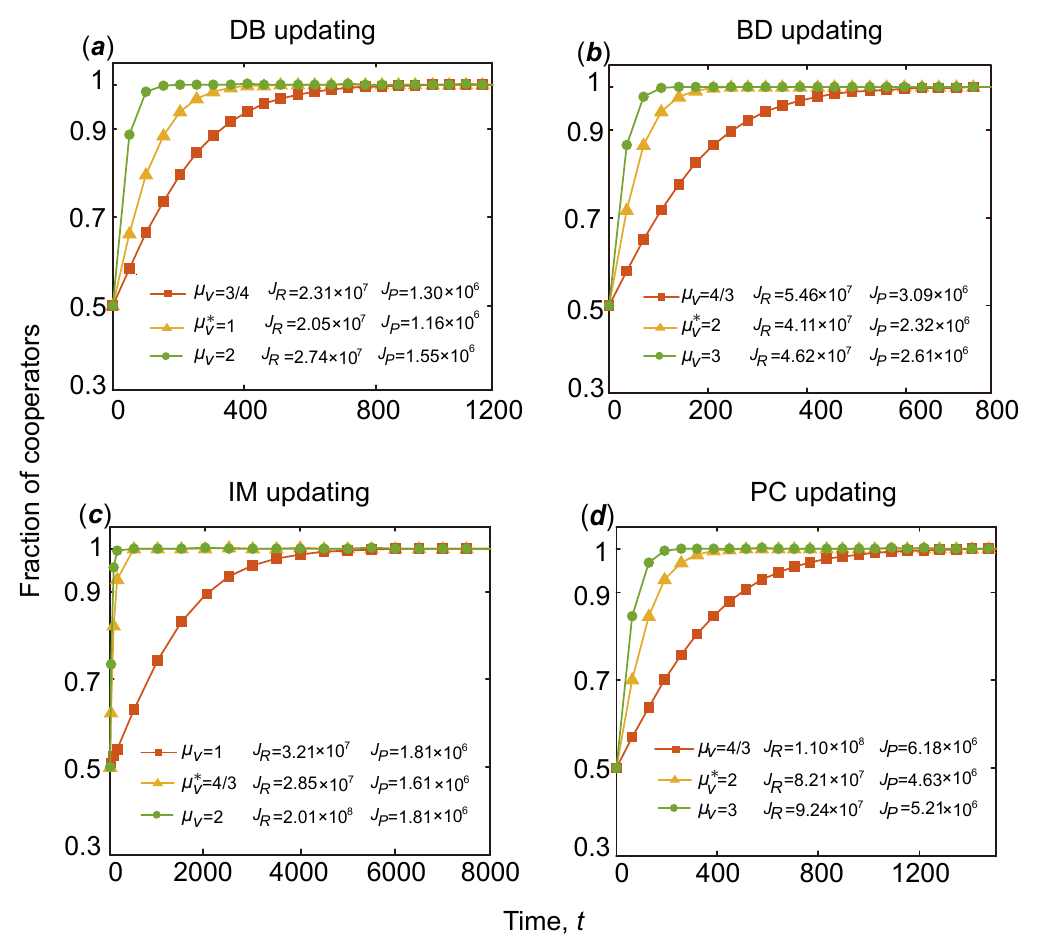}
\caption{\textbf{Time evolution of the fraction of cooperators for positive and negative incentives under different strategy update rules.}
Each panels shows the results derived from numerical calculations based on the obtained dynamical equation at different levels of incentives for reward ($R$) or punishment ($P$). The optimal incentive level is marked by {$\ast$}. For comparison we have also marked the $J_{R}$ and $J_{P}$ amounts of cumulative cost to reach the desired terminal state for each incentive protocol. Parameters: $N=100$, $b=2$, $c=1$, $\delta=0.01$, $\omega=0.01$, $p_{0}=0.5$, and $k=4$.}\label{fig3}
\end{figure}

In the following, we validate our analytical results for $\mu_{v}^{\ast}$ ($v\in \{R, P\}$) by means of numerical calculations and computer simulations. Figure~\ref{fig3} illustrates the fraction of cooperators $p_{C}$ as a function of time $t$ for the optimal $\mu^{\ast}_{v}$ and two other incentive schemes under four strategy update rule we considered. First, we note that our system always reaches the expected terminal state $p_{C}(t_{f})=0.99$ when we launch the evolution from a random $p_{0}=0.5$ state for each update rule. However, reaching this state requires significantly different costs for the incentive-providing institution, and this value is the lowest for $\mu_{v}^{\ast}$, no matter whether we apply reward or punishment. Notably, the optimal incentive protocol does not produce the fastest relaxation.

Moreover, we present the results of Monte Carlo simulations when positive or negative incentives are applied for each update rule (electronic supplementary material, figures S1-4). We can find that for different update rules the usage of optimal incentive protocol does not result in the fastest relaxation to the desired cooperation state, but it always leads to the smallest cumulative cost for the institution. We must stress that our findings are not limited to regular networks, but they remain valid for a broad range of interaction graphs from irregular random networks to small-world networks and scale-free networks. Our numerical calculations and simulations results coincide with the analytical predictions, and they illustrate that the cumulative cost is the lowest under the optimal incentive protocol.

\noindent\textbf{2.3. Comparison between optimal reward and punishment protocols}.

Through theoretical analysis presented in electronic supplementary material, we can conclude that the execution of optimal punishing scheme requires a lower cumulative cost, compared with the optimal rewarding scheme for each update rule, if the initial cooperation level is larger than the difference between the full and desired cooperation states. Otherwise, the usage of optimal rewarding scheme requires a lower cumulative cost. In order to verify such theoretical prediction and have an intuitive comparison, we show the cumulative cost values induced by the optimal punishing and rewarding protocols when the initial fraction $p_{0}$ of cooperators is adjustable. By assuming the optimal protocols of incentives, the requested cumulative cost can be determined by numerically integrating or by using Monte Carlo simulations.

\begin{figure}[ht]
\includegraphics[width=1\textwidth]{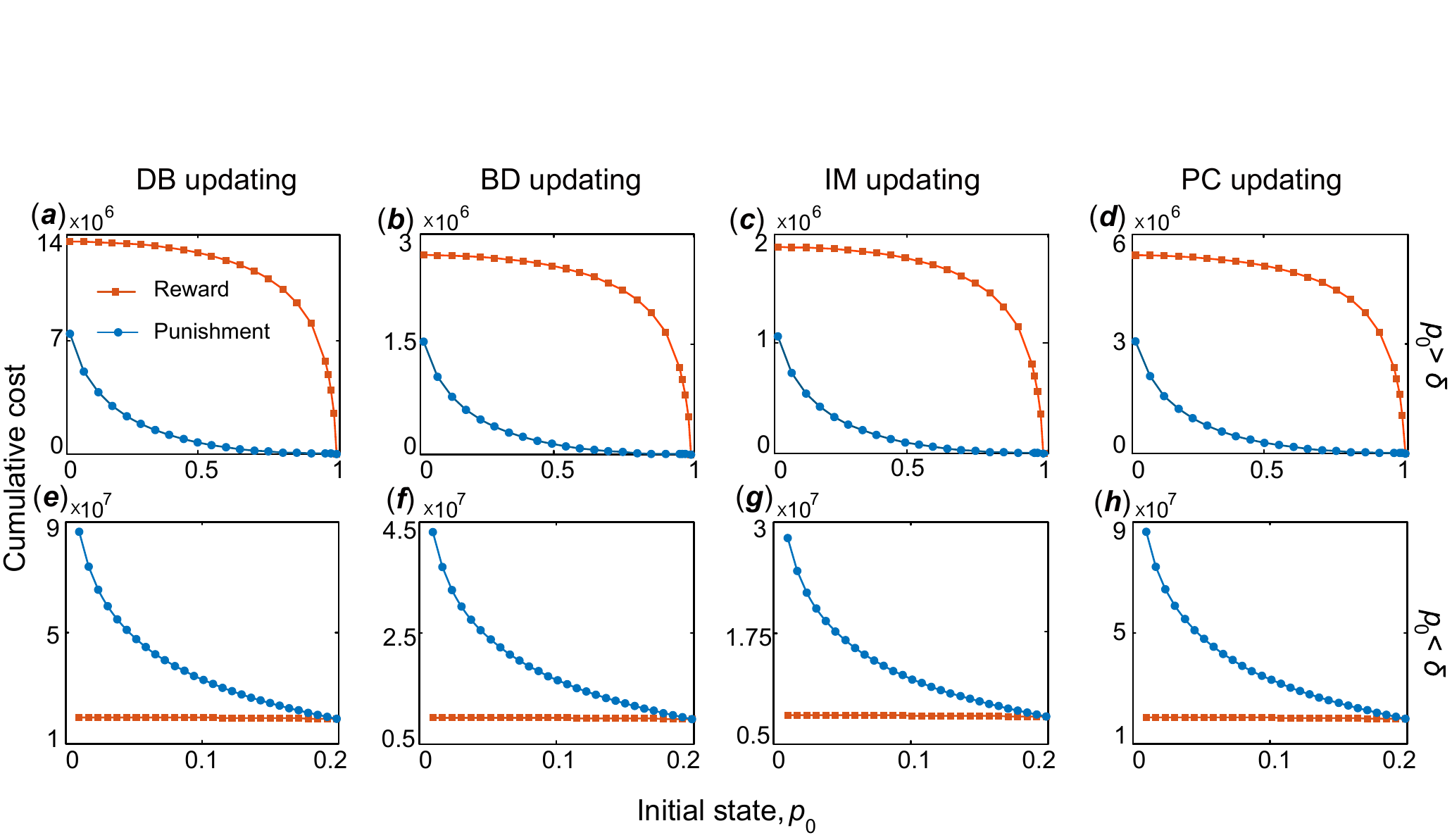}
\caption{\textbf{Cumulative cost needed for reaching the expected terminal state in dependence of the $p_{0}$ initial portion of cooperators for the optimal rewarding and punishing protocols for different strategy update rules.} Each column of panels represents a strategy update rule as indicated.
Top row represents the results of numerical calculations based on the obtained dynamical equation in the condition of $p_0>\delta=0.01$, while bottom row represents the results of numerical calculations in the condition of $p_0<\delta=0.2$. Other parameters: $N=100$, $b=2$, $c=1$, $\omega=0.01$, and $k=4$.}\label{fig4}
\end{figure}

Our numerical results are summarized in figure~\ref{fig4} for each update rule. From the top row of figure~\ref{fig4}, we observe that when the initial cooperation level is larger than the difference between the full cooperation and desired terminal states (with a set tolerance for defection, $\delta$), the institution needs to spend less cumulative cost to reach the expected cooperation state by means of punishing. On the contrary, from the bottom row of figure~\ref{fig4}, we see that when the initial cooperation level is less than the difference between the full cooperation and desired terminal states, the institution needs to spend less cumulative cost to reach the expected cooperation state by means of rewarding. However, in these different cases no matter whether the optimal rewarding or punishing protocol is applied, the corresponding cumulative cost value decreases as the initial cooperation level $p_{0}$ increases. In addition, our simulations results presented in figure S5 of electronic supplementary material also support our theoretical analysis, which provide a rigorous basis for the usage of incentives under different initial conditions in the context of evolutionary prisoner's dilemmas game in structured populations.

\vbox{}
\noindent\textbf{3. Discussion}\\
In human society, prosocial incentives are an essential means of avoiding
the ``tragedy of the commons'' \cite{Henrich_06,Gurerk_06,Ostrom90}. For incentive-providing institutions, however, the choice of incentives is mainly based on two aspects. One of them is knowing how much incentive is needed to promote the evolution of cooperation (potentially under different update rules) in structured populations. The other is whether the applied incentive scheme is an optimal time-dependent protocol requiring the minimal cost for the institution.

In order to investigate the above-mentioned tasks, we establish a game-theoretical framework. Namely, we consider the positive or negative incentive into the networked prisoner's dilemma game with four different strategy update rules, respectively. For a given update rule, we obtain the theoretical conditions of the minimal amounts of incentives needed for the evolution of cooperation. By establishing an index function for quantifying the executing cost, we derive the optimal positive and negative incentive protocols for each strategy update rule, respectively, by means of the approach of HJB equation. We find that these optimal incentive protocols are time-invariant for all the considered update rules. In addition, the optimal incentive protocols are identical both for negative and positive incentives. However, applying the punishing scheme requires a lower cumulative cost for the incentive-providing institution when the initial cooperation level is relatively high; otherwise, applying the rewarding scheme is cheaper. We further perform computer simulations, which confirm that our results are valid in different types of interaction topologies described by regular, random, small-world, and scale-free networks and thus demonstrate the general robustness of the findings.

In this work, we have quantified how much incentive is needed for the evolution of cooperation under each update rule we considered. However, when prosocial incentives are provided, the game structure may be changed. In particular, when the incentive amount $\mu_v>c$, the prisoner's dilemma game will be transformed into the harmony game, where cooperators dominate defectors naturally~\cite{Szab_07}. Interestingly, we note that if $\mu_v>c-b/k$ under DB updating and if $\mu_v>c-b/(k+2)$ under IM updating, cooperation is favored. This implies that under DB and IM rule, the $\mu_v$ value can be smaller than the cost $c$, which guarantees that the game structure is not changed. For BD and PC updating, however, the incentive amount needed for the evolution of cooperation must completely outweigh the cost of cooperation. Hence the effectiveness of interventions might shed some light on which update type can best capture population behaviors.

Here, we stress that the obtained optimal incentive protocols are time-invariant by solving the optimal control problems we formulated. More strikingly, the obtained optimal negative incentive protocol for each update rule is the same as the optimal positive incentive protocol. Thus, these optimal incentive protocols are state-independent. Then the institution does not need to monitor the population state from time to time for optimal  incentive implementation and can save monitoring costs since monitoring is generally costly. In addition, the optimal incentive level for DB and IM updating can guarantee that the dilemma faced by individuals is still a prisoner's dilemma. Accordingly, our work reflects that the incentive-based control protocols we obtained under DB and IM updating are simple and effective for promoting the evolution of cooperation in structured populations.

Incentives can be used as controlling tools to regulate the decision-making behaviors of individuals~\cite{Riehl_J18}. However, there are significant preference differences in the usage of punishment and reward for the evolution of cooperation from the perspectives of individuals and incentive-providing institutions \cite{Simon_12}. It has been suggested that punishment is often not preferred since the usage of punishment leads to a low total income or a low average payoff in repeated games \cite{David_09, Dreber_A08}. In contrast, incentive-providing institutions prefer to use punishments more frequently. This is because punishment incurs a lower cost of implementing incentives for the promotion of cooperation \cite{Simon_12, Sasaki_12}. Here, we consider the top-down-like incentive mechanism under which cooperators can be rewarded or defectors can be punished directly by the external centralized institution which has existed and works stably. In this framework, we strictly compare the rewarding and punishing schemes concerning implementation cost and show how the best choice depends on the initial cooperation level, providing a rigorous basis for the usage of incentives in the context of the prisoner's dilemma game and where it might apply in human society.

Extensions of our work are plentiful, especially in analytical terms. We would like to point out that we use the pair approximation approach to obtain our analytical results and this approach is used mainly for regular networks~\cite{ohtsuki_h06}. Indeed, this approach can be extended for other types of complex networks when the properties of these networks are additionally considered~\cite{Morita2008PTP,Overton2019JTB}. Along this line, hence it is worth analytically investigating the low-cost incentive policy for the evolution of cooperation by fully considering spatial properties of the underlying population structures. Furthermore, our theoretical analysis could be extended by resorting to methods relying on calculating the coalescence times of random walks~\cite{Allen_B17}. These methods might be particularly constructive when interventions at the topology level are at stake since they create a tighter link between the outcome and the network. In addition, we obtain our theoretical results in the limit of weak selection. Numerical analysis in other contexts has shown that due to relevant factors from the environment, the intensity of selection can change the game dynamics no matter the population is well-mixed or structured, and plays a crucial role in the determination of cost-efficient institutional incentive~\cite{pinheiro2012selection, Zisis2015sR, McAvoy2021ploscb, Han2018sr}. Hence, a natural question arising here is whether our theoretical results are still valid when selection is not weak~\cite{Ibsen-Jensen15PNAS}. Indeed, analytical calculations for strong selection remain tractable for some structures with high symmetry. Thus, there is potential to identify the theoretical conditions of how much incentive is needed to promote cooperation and to explore the optimal incentive protocols for these population structures.

In addition, our work focuses on minimizing the incentive costs up to a set level of cooperation in a population, irrespectively of how long that takes. Further research is required for situations in which the rate at which the transition happens can be important and a time-varying protocol might be the solution. We consider our research in the framework of the prisoner's dilemma game, which is a paradigm for studying the evolution of cooperation. There are other prototypical two-person dilemmas, e.g., snowdrift game~\cite{Doebeli05EL} and stag-hunt game~\cite{Skyrms04CUP}. A promising extension of this work is to consider these mentioned games for future study as well as group interactions or higher-order interactions~\cite{Perc13Interface,Li14SR,Li16PRE,Grilli17Nature,Alvarez-Rodriguez21NHB}. Furthermore, we design the cost-efficient incentive protocols by considering the global information (i.e., the fraction of cooperators in the whole population), but the local neighborhood properties on a structured network, such as how many cooperators are there in a neighbourhood, affect the final evolutionary outcomes~\cite{Lynch18IJCAI, Cimpeanu21KBS, Cimpeanu19ABMHuB}. Hence, it would be important to take into account these local information for optimal incentive protocols with minimal cost in the future work. Finally, other prosocial behaviours, such as honesty~\cite{Capraro20PRE} or trust and trustworthiness~\cite{Kumar20Interface}, are also fundamental for cooperation, and hence it is a meaningful extension to study the optimization problems of incentives for promoting the evolution of these behaviors.

%%%%Capraro20PRE, Kumar20Interface,

\vbox{}
\noindent\textbf{4. Model and Methods}

\noindent \textbf{4.1. Prisoner's dilemma game}

We consider that a population of individuals are distributed on the nodes of an interaction graph. At each round, each individual plays the evolutionary prisoner's dilemma game with its neighbors and can choose to cooperate ($C$) or defect ($D$). We consider the payoff matrix for the game as
\begin{eqnarray}
\bordermatrix{
      &C  &D  \cr
C &b-c & -c\cr
D &b & 0 },
\label{matrix}
\end{eqnarray}
where $b$ represents the benefit of cooperation and $c$ ($0<c<b$) represents the cost of cooperation. After engaging in the pairwise interactions with all the adjacent neighbors, each individual collects its payoff based on the payoff matrix.

\noindent \textbf{4.2. Institutional incentives}

Furthermore, prosocial incentives provided by incentive-providing institutions can be used to reward cooperators or punish defectors after they play the game with their neighbors. Here, we consider both types of incentives, i.e., positive and negative incentives, respectively. If positive incentives are used, then a cooperator in the game is rewarded with a $\mu_{R}$ amount received from a central institution when interacting with a neighbor \cite{Riehl_J18, David_09}. If negative incentives are used, then a defector is fined by a $\mu_{P}$ amount for each interaction \cite{Dreber_A08}. Consequently, when positive incentives are used, the modified payoff matrix becomes
\begin{eqnarray}
\bordermatrix{
      &C  &D  \cr
C &b-c+\mu_{R} & -c+\mu_{R}\cr
D &b& 0 },
\end{eqnarray}
and when negative incentives are used, the modified payoff matrix becomes
\begin{eqnarray}
\bordermatrix{
      &C  &D  \cr
C &b-c & -c\cr
D &b-\mu_{P} & -\mu_{P} }.
\end{eqnarray}
As a result, based on the above payoff matrices each individual collects its total payoff, which is derived from the pairwise interactions with neighbors and the incentive-providing institution.

\noindent\textbf{4.3. Strategy update rules}

According to the evolutionary selection principle, players update their strategies from time to time, but the way how to do it may influence the evolutionary outcome significantly \cite{Szab_07,ohtsuki_h06}. In agreement with previous works~\cite{ohtsuki_h06,ohtsuki_1}, we here consider four major strategy update rules, describing DB, BD, IM, and PC updating. Specifically, for DB updating, at each time step a random individual from the entire population is chosen to die; subsequently the neighbors compete for the empty site with probability proportional to their fitness. For BD updating, at each time step an individual is chosen for reproduction from the entire population with probability  proportional to fitness; the offspring of this individual replaces a randomly selected neighbor. For IM updating, at each time step a random individual from the entire population is chosen to update its strategy; it will either stay with its own strategy or imitate one of the neighbors' strategies with probability  proportional to their fitness. For PC updating, at each time step a random individual is chosen to update its strategy, and it compares its own fitness with a randomly chosen neighbor. The focal individual either keeps its current strategy or adopts the neighbor's strategy with a probability that depends on the fitness difference.

\noindent \textbf{4.4. Optimazing incentives}

Since providing incentives is costly for institutions, it has a paramount importance to find the optimal $\mu^*_{v}$, which requires the minimal effort but is still capable of supporting cooperation effectively. To reach this goal, we first establish an index function for quantifying the executing cumulative cost, which is expressed as
 \begin{eqnarray}
J_{v}=\int^{t_{f}}_{t_0}\frac{(kNp_{i}\mu_{v})^{2}}{2}dt,
 \end{eqnarray}
where $p_{i}=p_{C}$ if $v=R$ which means that positive incentives are applied, otherwise $p_{i}=p_{D}$ which means that negative incentives are applied. Here $t_{0}$ is the initial time and is set to $0$ in this work, and $t_{f}$ the terminal time for the system. Based on the above description, we then explore the optimal incentive protocol during the evolutionary period between $0$ and $t_{f}$ by using optimal control theory \cite{Evans_05, Geering_07, Lenhart_05}. It is a crucial assumption that $t_{f}$ is not fixed, but we monitor the evolution until the fraction of cooperators reaches the target level $p_{C}(t_{f})$  at $t_{f}$. Here, we suppose that $p_{C}(t_{f})=1-\delta>p_{0}$, where $p_{0}$ ($p_{0}>0$) is the initial cooperation level and $\delta$ is the parameter determining the expected cooperation level at $t_{f}$, satisfying $0\leq \delta<1-p_0$.

Accordingly, we formulate the optimal control problem for reward or punishment given as
\begin{eqnarray}
&\min\,\,&J_{v}=\int^{t_{f}}_{0}\frac{(kNp_{i}\mu_{v})^{2}}{2}dt,\nonumber\\
&{ s.t.}&\quad  \left\{\begin{array}{lc}
\frac{dp_{C}}{dt}=F_{j}(p_{C}, \mu_{v}, t), j=\textrm{DB, BD, IM, or PC},\\
p_{C}(0)=p_{0}, \\
p_{C}(t_{f})=1-\delta.
\end{array}\right.
\end{eqnarray}

Here the cost function $J_v$ characterizes the cumulative cost on average during the period $[0, t_{f}]$ for the dynamical system to reach the terminal state $1-\delta$ from the initial state $p_{0}$. Thus the quantity $\min J_v$ can work as the objective of calculating the optimal incentive protocol $\mu^*_{v}$ with the minimal executing cost. The details of solving the optimal control problems can be found in sections~1-4 of electronic supplementary material.

\noindent \textbf{4.5. Monte Carlo simulations}

During a full Monte Carlo step, on average each player has a chance to update its strategy. The applied four different update rules are specified above. Besides, we have tested alternative interaction topologies, including regular networks generated by using a two-dimensional square lattice of size $N=L\times L$ \cite{Nowak_92} and scale-free networks obtained by using preferential-attachment model \cite{Barab_99} starting from $m_{0}=6$ where at every time step each new node is connected to $m=2$ existing nodes fulfilling the standard power-law distribution. Alternatively, Erd\H{o}s-R\'{e}nyi random graph model \cite{Erd_59} and small-world networks of Watts-Strogatz model with $p=0.1$ rewiring parameter \cite{Watts_98} are considered. These simulation results are summarized in figures S1-5 of electronic supplementary material.

\noindent \\ \textbf{Author contributions} \\
X.C., S.W., V.V.V., and A.S. designed the research, S.W. and Z.X. performed the research, X.C., V.V.V., A.S., and S.W. wrote the manuscript, and all authors discussed the results and commented on and improved the manuscript.

\noindent \\ \textbf{Data Accessibility} \\
This article has no additional data.

\noindent \\ \textbf{Competing financial interests} \\
The authors declare no competing financial interests.

\noindent \\ \textbf{Ethics} \\
This article does not present research with ethical considerations.

\noindent \\ \textbf{Funding} \\
This research was supported by the National Natural Science Foundation of China (Grant Nos. 61976048 and 62036002) and the Fundamental Research Funds of the Central Universities of China. S.W. acknowledges the support from China Scholarship Council (Grant No. 202006070122).  A.S. was supported by the National Research, Development and Innovation Office (NKFIH) under Grant No. K142948. V.V.V. acknowledges funding from the Computational Science Lab - Informatics Institute of the University of Amsterdam.

\vbox{}

\clearpage

\clearpage

\appendix

\centerline{Electronic Supplementary Material for}
\vbox{}

\centerline{\textbf{Optimization of institutional incentives for cooperation in structured populations}}
\vbox{}

\centerline{Shengxian Wang, Xiaojie Chen, Zhilong Xiao, Attila Szolnoki, and V\'{\i}tor V. Vasconcelos}

\vbox{}

In electronic supplementary material, we provide a detailed theoretical analysis to explore optimal incentive protocols for the promotion of cooperation in structured populations. Specifically,  we consider four different strategy update rules, describing DB updating in section 1, BD updating in section 2, IM updating in section 3, and finally PC updating in section 4. In each section, we first use the pair approximation method to explore the dynamical equation with positive or negative incentive and theoretically obtain the conditions of the minimal amounts of incentives needed for the evolution of cooperation. After that, we formulate optimal control problems for both positive and negative incentive protocols and obtain the optimal positive and negative incentive protocols by means of the approach of HJB equation. As a result, the requested cumulative costs to reach the expected final state are determined for the optimal incentive protocols.

\vbox{}
\leftline{\textbf{1. \,\, DB Updating}}
\noindent \textbf{1.1. \,\, Positive Incentive}\\
Population structure is represented by a regular network of $N$ nodes with degree $k>2$. The vertices of network correspond to individuals and the edges represent who interacts with whom. Each individual plays the Prisoner's Dilemma game with its neighbors, who can either cooperate ($C$) or defect ($D$). Here, we introduce some notations. Let $p_{i}$ denote the proportion of individuals with strategy $i$, let $p_{ij}$ denote the  proportion of $ij$--pairs, and finally let $q_{i|j}$  denote the conditional probability of finding an $i$--individual given that the neighboring node is a $j$--individual, where $i,j \in \{C, D\}$. By using these notations, we have that $p_{C}+p_{D}=1$, $q_{C|i}+q_{D|i}=1$, $p_{ij}=q_{i|j}p_{j}$, and $p_{CD}=p_{DC}$.

We first consider the positive incentive into the networked Prisoner's Dilemma game with DB updating~\cite{ohtsuki_h06, ohtsuki_1}. According to DB updating, we randomly select a focal individual to die with probability $p_{i}$ ($i\in \{C, D\}$), where $i$ represents the strategy of the focal individual. Let $k_{C}$ and $k_{D}$ denote the numbers of cooperators and defectors among its $k$ neighbors with $k_{D}+k_{C}=k$. If the focal individual adopts strategy $D$, then the fitness of a $C$--neighbor is
\begin{equation}
\begin{aligned}
f_{C} = 1-\omega+\omega\{(b-c+\mu_{R})(k-1)q_{C\mid C}+(\mu_{R}-c)[(k-1)q_{D\mid C}+1]\},
\end{aligned}\tag{S1}
\end{equation}
and the fitness of a $D$--neighbor is
\begin{equation}
\begin{aligned}
f_{D}=1-\omega+\omega[b(k-1)q_{C\mid D}],
\end{aligned}\tag{S2}
\end{equation}
where  $0\leq\omega\leq1$ measures the strength of selection.

Since all the neighbors of the focal individual compete for the empty site with probability  proportional to their fitness, the probability that a $C$--neighbor replaces this empty site is given by
\begin{equation}
\begin{aligned}
\Gamma=\frac{k_{C}f_{C}}{k_{C}f_{C}+k_{D}f_{D}}.
\end{aligned}\tag{S3}
\end{equation}
Based on the above equations, $p_{C}$ increases by $1/N$ with probability
\begin{equation}
\begin{aligned}
P(\Delta p_{C}=\frac{1}{N})=p_{D}\sum_{k_{C}=0}^{k} \binom{k}{k_{C}}(q_{C|D})^{k_{C}}(q_{D|D})^{k_{D}}\Gamma,
\end{aligned}\tag{S4}
\end{equation}
and the number of $CC$-pairs increases by $k_{C}$ and hence $p_{CC}$ increases by $k_{C}/(kN/2)$ with probability
\begin{equation}
\begin{aligned}
P(\Delta p_{CC}=\frac{2k_{C}}{kN})=p_{D}\binom{k}{k_{C}} (q_{C|D})^{k_{C}}(q_{D|D})^{k_{D}}\Gamma.
\end{aligned}\tag{S5}
\end{equation}

Furthermore, we consider another case, i.e., the randomly selected focal individual adopts strategy $C$. In this case, the fitness of a $C$--neighbor is
\begin{equation}
\begin{aligned}
f_{C}=1-\omega+ \omega\{(b-c+\mu_{R})[(k-1)q_{C|C}+1]+(\mu_{R}-c)(k-1)q_{D|C}\},
\end{aligned}\tag{S6}
\end{equation}
and the fitness of a $D$--neighbor is
\begin{equation}
\begin{aligned}
f_{D}=1-\omega+ \omega\{b[(k-1)q_{C|D}+1]+0\cdot(k-1)q_{D|D}\}.
\end{aligned}\tag{S7}
\end{equation}
The probability that a $D$--neighbor replaces the empty site is given by
\begin{equation}
\begin{aligned}
M=\frac{k_{D}f_{D}}{k_{C}f_{C}+k_{D}f_{D}}.
\end{aligned}\tag{S8}
\end{equation}
Therefore, $p_{C}$ decreases by $1/N$ with probability
\begin{equation}
\begin{aligned}
P(\Delta p_{C}=-\frac{1}{N})=p_{C}\sum_{k_{C}=0}^{k} \binom{k}{k_{C}}(q_{C|D})^{k_{C}}(q_{D|D})^{k_{D}}M,
\end{aligned}\tag{S9}
\end{equation}
and the number of $CC$--pairs decreases by $k_{C}$ and hence $p_{CC}$ decreases by $k_{C}/(kN/2)$ with probability
\begin{equation}
\begin{aligned}
 P(\Delta p_{CC}=-\frac{2k_{C}}{kN})= p_{C}\binom{k}{k_{C}}(q_{C|D})^{k_{C}}(q_{D|D})^{k_{D}}M.
\end{aligned}\tag{S10}
\end{equation}
We suppose that one replacement event occurs in one unit of time, and the derivative of
$p_{C}$ can be written as
\begin{equation}
\begin{aligned}
\frac{dp_{C}}{dt}&=\frac{E(\Delta p_{C})}{\Delta t}=\frac{\frac{1}{N}P(\Delta p_{C}=\frac{1}{N})-\frac{1}{N}P(\Delta p_{C}=-\frac{1}{N})}{\frac{1}{N}} \nonumber\\
&=\omega \frac{k-1}{k}p_{CD}[(\mu_{R}-c)+(k-1)\eta_{1}](q_{C\mid C}+q_{D\mid D})+o(\omega^{2}),
\end{aligned}\tag{S11}
\end{equation}
in which $\eta_{1}=(b-c+\mu_{R})q_{C\mid C}+(\mu_{R}-c)q_{D\mid C}-bq_{C\mid D}$.
Accordingly, the derivative of $p_{CC}$ is given by
\begin{equation}
\begin{aligned}
\frac{dp_{CC}}{dt}&=\frac{E(\Delta p_{CC})}{\Delta t}= \frac{\sum_{k_{C}=0}^k \frac{2k_{C}}{kN}P(\Delta p_{CC}=\frac{2k_{C}}{kN})- \sum_{k_{C}=0}^k \frac{2k_{C}}{kN}P(\Delta p_{CC}=-\frac{2k_{C}}{kN})}{\frac{1}{N}} \nonumber\\
 &=\frac{2p_{CD}}{k}[1+(k-1)(q_{C\mid D}-q_{C\mid C})]+o(\omega).
\end{aligned}\tag{S12}
\end{equation}
Due to $q_{C\mid C}=\frac{p_{CC}}{p_{C}}$, we have
\begin{equation}
\begin{aligned}
\frac{dq_{C|C}}{dt}=\frac{d}{dt}(\frac{p_{CC}}{p_{C}})=\frac{2 p_{CD}}{k p_{C}}[1+(k-1)(q_{C\mid D}-q_{C\mid C})]+o(\omega).
\end{aligned}\tag{S13}
\end{equation}
Other variables, such as $p_{D}$ and $q_{D\mid C}$, can also be expressed by $p_{C}$ and $q_{C|C}$ through appropriate calculation, and then the dynamical system can be described by $p_{C}$ and $q_{C|C}$. Rewriting the right-hand expressions of Eqs.~(S11) and~(S13) as functions of $p_{C}$ and $q_{C|C}$ yields the dynamical equation given by
\begin{equation}
\begin{aligned}
\left\{\begin{array}{lc}
\frac{dp_{C}}{dt}=\omega \Psi_{\textrm{DB}}^R(p_{C}, q_{C\mid C})+o(\omega^{2}),\nonumber\\
\frac{dq_{C|C}}{dt}=\Phi_{\textrm{DB}}^R(p_{C}, q_{C\mid C})+o(\omega),
\end{array}\right.
\end{aligned}\tag{S14}
\end{equation}
where
\begin{equation*}
\begin{aligned}
\left\{\begin{array}{lc}
 \Psi_{\textrm{DB}}^{R}(p_{C}, q_{C\mid C})=\frac{k-1}{k}p_{CD}[(-c+\mu_{R})+(k-1)\eta_{1}](q_{C\mid C}+q_{D\mid D}),\nonumber\\
  \Phi_{\textrm{DB}}^{R}(p_{C}, q_{C\mid C})=\frac{2p_{CD}}{k p_{C}}[1+(k-1)(q_{C\mid D}-q_{C\mid C})].
\end{array}\right.
\end{aligned}\tag{S15}
\end{equation*}
Under weak selection ($w\ll 1$), the velocity of $q_{C|C}$ can be large, and it may rapidly converge to the root defined by $\Phi_{\textrm{DB}}^{R}(p_{C}, q_{C\mid C})=0$ as time $t\rightarrow +\infty$.
Thus, we get
\begin{equation}
\begin{aligned}
  q_{C|C}=p_{C}+\frac{1}{k-1}(1-p_{C}).
\end{aligned}\tag{S16}
\end{equation}
Accordingly, the dynamical equation described by Eq.~(S14) becomes
 \begin{equation}
\begin{aligned}
\frac{dp_{C}}{dt}=\frac{\omega(k-2)[b+k(\mu_{R}-c)]}{k-1}p_{C}(1-p_{C})+o(\omega^{2}),
\end{aligned}\tag{S17}
\end{equation}
which has two fixed points $p_{C}=0$ and $p_{C}=1$. We define the function $F_{\textrm{DB}}(p_{C}, \mu_R, t)$ as
\begin{equation}
\begin{aligned}
F_{\textrm{DB}}(p_{C}, \mu_R, t)=\frac{\omega(k-2)[b+k(\mu_{R}-c)]}{k-1}p_{C}(1-p_{C})+o(\omega^{2}).
\end{aligned}\tag{S18}
\end{equation}
This function is a continuously differentiable function, and the derivative of $F_{\textrm{DB}}(p_{C}, \mu_R, t)$ with respect to $p_{C}$ is
\begin{equation}
\begin{aligned}
\frac{dF_{\textrm{DB}}}{dp_{C}}=\frac{\omega(k-2)[b+k(\mu_{R}-c)]}{k-1}(1-2p_{C})+o(\omega^{2}).
\end{aligned}\tag{S19}
\end{equation}
For $\mu_{R}>c-\frac{b}{k}$, we have $\frac{dF_{\textrm{DB}}}{dp_{C}}|_{{p}_{C}=1}=-\frac{\omega(k-2)[b+k(\mu_{R}-c)]}{k-1}<0$ and $\frac{dF_{\textrm{DB}}}{dp_{C}}|_{{p}_{C}=0}=\frac{\omega(k-2)[b+k(\mu_{R}-c)]}{k-1}>0$. This means that the fixed point $p_{C}=1$ is stable and $p_{C}=0$ unstable, i.e., cooperators prevail over defectors.

Lastly, we then study the special case of $\mu_{R}=0$. In this case, we can see that for $b/c>k$,  the fixed point $p_{C}=1$ is stable and  $p_{C}=0$ unstable. Thus, we obtain the condition $b/c>k$ for the evolution of cooperation as previously obtained in Refs.~\cite{ohtsuki_h06, ohtsuki_1}.

\vbox{}
\noindent \textbf{1.2. \,\, Negative Incentive}\\
In this subsection, we then consider the negative incentive into the networked Prisoner's Dilemma game with DB updating, and the payoff matrix is given by Eq.~(15) in the main text. According to DB updating, if the focal individual adopts strategy $D$, then the fitness of a $C$--neighbor is
\begin{equation}
\begin{aligned}
f_{C}=1-\omega+\omega\{(b-c)(k-1)q_{C\mid C}-c[(k-1)q_{D\mid C}+1]\},
\end{aligned}\tag{S20}
\end{equation}
and the fitness of a $D$--neighbor is
\begin{equation}
\begin{aligned}
f_{D}=1-\omega+\omega\{(b-\mu_{P})(k-1)q_{C\mid D}-\mu_{P}[(k-1)q_{D\mid D}+1]\}.
\end{aligned}\tag{S21}
\end{equation}
The probability that a $C$--neighbor replaces the empty site is given by the expression $\Gamma$ in Eq.~(S3).  Therefore, $p_{C}$ increases by $1/N$ with probability
\begin{equation}
\begin{aligned}
P(\Delta p_{C}=\frac{1}{N})=p_{D}\sum_{k_{C}=0}^{k}\binom{k}{k_{C}}(q_{C|D})^{k_{C}}(q_{D|D})^{k_{D}}\Gamma.
\end{aligned}\tag{S22}
\end{equation}
Accordingly, the number of $CC$--pairs increases by $k_{C}$ and hence $p_{CC}$ increases by $k_{C}/(kN/2)$ with probability
\begin{equation}
\begin{aligned}
P(\Delta p_{CC}=\frac{2k_{C}}{kN})=p_{D}\binom{k}{k_{C}} (q_{C|D})^{k_{C}}(q_{D|D})^{k_{D}}\Gamma.
\end{aligned}\tag{S23}
\end{equation}

In addition, we consider another case where the focal individual adopts strategy $D$. In this case, the fitness of a $C$--neighbor is
\begin{equation}
\begin{aligned}
f_{C}=1-\omega+\omega\{(b-c)[(k-1)q_{C|C}+1]-c(k-1)q_{D|C}\},
\end{aligned}\tag{S24}
\end{equation}
and  the fitness of a $D$--neighbor is
\begin{equation}
\begin{aligned}
f_{D}=1-\omega+\omega\{(b-\mu_{P})[(k-1)q_{C|D}+1]-\mu_{P}(k-1)q_{D|D}\}.
\end{aligned}\tag{S25}
\end{equation}
The probability that a $D$--neighbor replaces the empty site can be also given by the expression $M$ in Eq.~(S8). Thus, $p_{C}$ decreases by $1/N$ with probability
\begin{equation}
\begin{aligned}
 P(\Delta p_{C}=-\frac{1}{N})= p_{C}\sum_{k_{C}=0}^{k}\binom{k}{k_{C}}(q_{C|D})^{k_{C}}(q_{D|D})^{k_{D}}M.
\end{aligned}\tag{S26}
\end{equation}
Accordingly, the number of $CC$--pairs decreases by $k_{C}$ and $p_{CC}$ decreases by $k_{C}/(kN/2)$ with probability
\begin{equation}
\begin{aligned}
 P(\Delta p_{CC}=-\frac{2k_{C}}{kN})=p_{C}\binom{k}{k_{C}} (q_{C|D})^{k_{C}}(q_{D|D})^{k_{D}}M.
\end{aligned}\tag{S27}
\end{equation}
Based on these calculations, we obtain the time derivative of
$p_{C}$ given by
\begin{equation}
\begin{aligned}
\frac{dp_{C}}{dt}&=\frac{E(\Delta p_{C})}{\Delta t}=\frac{\frac{1}{N}P(\Delta p_{C}=\frac{1}{N})-\frac{1}{N}P(\Delta p_{C}=-\frac{1}{N})}{\frac{1}{N}} \nonumber\\
 &=\frac{\omega(k-1)}{k}p_{CD}[\mu_{P}-c+(k-1)\eta_{2}](q_{C\mid C}+q_{D\mid D})+o(\omega^{2}),
\end{aligned}\tag{S28}
\end{equation}
in which $\eta_{2}=(b-c)q_{C\mid C}-c q_{D\mid C}+(\mu_{P}-b)q_{C\mid D}+\mu_{P}q_{D|D}$.
And the time derivative of $p_{CC}$ is given by
\begin{equation}
\begin{aligned}
\frac{dp_{CC}}{dt}&=\frac{E(\Delta p_{CC})}{\Delta t}= \frac{\sum_{k_{C}=0}^k \frac{2k_{C}}{kN}P(\Delta p_{CC}=\frac{2k_{C}}{kN})- \sum_{k_{C}=0}^k \frac{2k_{C}}{kN}P(\Delta p_{CC}=-\frac{2k_{C}}{kN})}{\frac{1}{N}} \nonumber\\
 &=\frac{2p_{CD}}{k}[1+(k-1)(q_{C\mid D}-q_{C\mid C})]+o(\omega).
\end{aligned}\tag{S29}
\end{equation}
Furthermore, we have
\begin{equation}
\begin{aligned}
\frac{dq_{C|C}}{dt}=\frac{2p_{CD}}{k p_{C}}[1+(k-1)(q_{C\mid D}-q_{C\mid C})]+o(\omega).
\end{aligned}\tag{S30}
\end{equation}
Hence, the dynamical equation can be described by
\begin{equation}
\begin{aligned}
\left\{\begin{array}{lc}
\frac{dp_{C}}{dt}=\omega \Psi_{\textrm{DB}}^P(p_{C}, q_{C\mid C})+o(\omega^{2}),\nonumber\\
\frac{dq_{C|C}}{dt}=\Phi_{\textrm{DB}}^P(p_{C}, q_{C\mid C})+o(\omega),
 \end{array}\right.
 \end{aligned}\tag{S31}
\end{equation}
where \begin{equation*}
\begin{aligned}
\left\{\begin{array}{lc}
        \Psi_{\textrm{DB}}^P(p_{C}, q_{C\mid C})=\frac{k-1}{k}p_{CD}[\mu_{P}-c+(k-1)\eta_{2}](q_{C\mid C}+q_{D\mid D}), \\
       \Phi_{\textrm{DB}}^P(p_{C}, q_{C\mid C})=\frac{2p_{CD}}{k p_{C}}[1+(k-1)(q_{C\mid D}-q_{C\mid C})].
      \end{array}\right.
 \end{aligned}
      \end{equation*}
Under weak selection, the velocity of $q_{C|C}$ can be large, and it may rapidly converge to the root defined by $\Phi_{\textrm{DB}}^P(p_{C}, q_{C\mid C})=0$ as time $t\rightarrow +\infty$.
Thus, we get
\begin{equation}
\begin{aligned}
 q_{C|C}=p_{C}+\frac{1}{k-1}(1-p_{C}).
\end{aligned}\tag{S32}
\end{equation}
Correspondingly, the dynamical equation described by Eq.~(S31) becomes
\begin{equation}
\begin{aligned}
\frac{dp_{C}}{dt}=\frac{\omega(k-2)[b+k(\mu_{P}-c)]}{k-1}p_{C}(1-p_{C})+o(\omega^{2}),
\end{aligned}\tag{S33}
\end{equation}
which has two fixed points $p_{C}=0$ and $p_{C}=1$. We define the function $F_{\textrm{DB}}(p_{C}, \mu_P, t)$ as
\begin{equation}
\begin{aligned}
F_{\textrm{DB}}(p_{C}, \mu_P, t)=\frac{\omega(k-2)[b+k(\mu_{P}-c)]}{k-1}p_{C}(1-p_{C})+o(\omega^{2}),
\end{aligned}\tag{S34}
\end{equation}
and the derivative of $F_{\textrm{DB}}(p_{C}, \mu_P, t)$  with respect to $p_{C}$ is
\begin{equation}
\begin{aligned}
\frac{dF_{\textrm{DB}}}{dp_{C}}=\frac{\omega(k-2)[b+k(\mu_{P}-c)]}{k-1}(1-2p_{C})+o(\omega^{2}).
\end{aligned}\tag{S35}
\end{equation}
Hence, for $\mu_{P}>c-\frac{b}{k}$ we have $\frac{dF_{\textrm{DB}}}{dp_{C}}|_{{p}_{C}=1}=-\frac{\omega(k-2)[b+k(\mu_{P}-c)]}{k-1}<0$ and $\frac{dF_{\textrm{DB}}}{dp_{C}}|_{{p}_{C}=0}=\frac{\omega(k-2)[b+k(\mu_{P}-c)]}{k-1}>0$, which means that the fixed point $p_{C}=1$ is stable and $p_{C}=0$ unstable, i.e.,  cooperators prevail over defectors. Particularly, when $\mu_{P}=0$,  we can see that for $b/c>k$, the fixed point $p_{C}=1$ is stable and $p_{C}=0$ unstable. Thus, we obtain the condition $b/c>k$ for the evolution of cooperation as obtained in Refs.~\cite{ohtsuki_h06, ohtsuki_1}.

\vbox{}
\noindent \textbf{1.3. \,\, Optimal Incentive Protocols}\\
In subsections 1.1 and 1.2,  we have theoretically derived the dynamical system with positive or negative incentive by means of the pair approximation method in the limit of weak selection, which is given by
\begin{equation}
\begin{aligned}
\frac{dp_{C}}{dt}=F_{\textrm{DB}}(p_{C}, \mu_{v}, t)=\frac{\omega(k-2)[b+k(\mu_{v}-c)]}{k-1}p_{C}(1-p_{C})+o(\omega^{2}).
\end{aligned}\tag{S36}
\end{equation}
As noted, this dynamical system has two equilibria which are $p_{C}=0$ and $p_{C}=1$. If $\mu_{v}>c-\frac{b}{k}$, the former is unstable and the latter is stable, which means that cooperation will be promoted in the long run. Since providing incentive is costly, our principal goal is to explore the optimal incentive protocol that is still able not only to promote cooperation, but also requires a minimal cost. To do that, we now solve the formulated optimal control problems for DB updating.

First, we solve the optimal control problem for rewarding. The Hamiltonian function $H_{\textrm{DB}}(p_{C}, \mu_{R}, t)$ is defined as
\begin{equation}
\begin{aligned}
H_{\textrm{DB}}(p_{C}, \mu_{R}, t)=\frac{(kNp_{C}\mu_{R})^{2}}{2}+\frac{\partial J_{R}^{\ast}}{\partial p_{C}}F_{\textrm{DB}}(p_{C}, \mu_{R}, t),
\end{aligned}\tag{S37}
\end{equation}
where $J_{R}^\ast$ is the optimal cost function of $p_{C}$ and $t$ for the optimal rewarding protocol, given as
\begin{equation}
\begin{aligned}
J_{R}^\ast=\int^{t_{f}}_{0}\frac{(kNp_{C}\mu_{R}^\ast)^{2}}{2}dt.
\end{aligned}\tag{S38}
\end{equation}
By solving $\frac{\partial H_{\textrm{DB}}}{\partial \mu_{R}}=0$, we know that the optimal rewarding protocol $\mu_{R}^{\ast}$ should satisfy
\begin{equation}
\begin{aligned}
\mu_{R}^{\ast}=-\frac{\omega (k-2)(1-p_{C})}{N^{2}k(k-1)p_{C}}\frac{\partial J_{R}^{\ast}}{\partial p_{C}}.
\end{aligned}\tag{S39}
\end{equation}
Generally, we should solve the canonical equations of Eq.~(S37) to obtain the optimal rewarding protocol \cite{Evans_05, Geering_07, Lenhart_05}. Yet, the obtained dynamical systems are nonlinear which greatly increases the complexity of obtaining the exact expression of the optimal protocols by a direct calculation. Instead,  to solve the optimal control problem we use the dynamic programming method, HJB equation for continuous-time systems \cite{Evans_05, Geering_07, Lenhart_05}. This equation can be written as
\begin{equation}
\begin{aligned}
-\frac{\partial J_{R}^\ast}{\partial t}=H_{\textrm{DB}}(p_{C}, \mu_{R}^{\ast}, t).
\end{aligned}\tag{S40}
\end{equation}
By substituting Eq.~(S39) into the above HJB equation, we have
\begin{equation}
\begin{aligned}
-\frac{\partial J_{R}^{\ast}}{\partial t}= \frac{(kNp_{C}\mu_{R}^\ast)^{2}}{2}+\frac{\omega (k-2)[b+k(\mu^{\ast}_{R}-c)]}{k-1}p_{C}(1-p_{C})\frac{\partial J_{R}^{\ast}}{\partial p_{C}}.
\end{aligned}\tag{S41}
\end{equation}
Since we assume that the terminal time $t_{f}$ is not fixed, the optimal cost function $J_{R}^{\ast}(p_{C},t)$ is independent of $t$. Consequently, we have
\begin{equation}
\begin{aligned}
 \frac{\partial J_{R}^\ast}{\partial t}=0.
\end{aligned}\tag{S42}
\end{equation}
We then yield
\begin{equation}
\begin{aligned}
 \frac{\partial J_{R}^{\ast}}{\partial p_{C}} = 0 \;\; {\rm or} \;\;
\frac{\partial J_{R}^{\ast}}{\partial p_{C}}=\frac{2 N^{2}(k-1)(b-ck)p_{C}}{\omega (k-2)(1-p_{C})}.
\end{aligned}\tag{S43}
\end{equation}
As $\mu_{R}>0$ and $p_{C}\in(0, 1)$, we have
\begin{equation}
\begin{aligned}
\frac{\partial J_{R}^{\ast}}{\partial p_{C}}<0.
\end{aligned}\tag{S44}
\end{equation}
From Eq.~(S44), we can see that this inequality is obviously satisfied for $b/c\geq k$. Instead, we consider the case, i.e., $b/c<k$,  and hence only $\frac{\partial J_{R}^{\ast}}{\partial p_{C}}=\frac{2 N^{2}(k-1)(b-ck)p_{C}}{\omega (k-2)(1-p_{C})}$ holds. By substituting this equation into Eq.~(S39), we obtain the optimal rewarding level as
\begin{equation}
\begin{aligned}
\mu_{R}^{\ast}=\frac{2(ck-b)}{k}.
\end{aligned}\tag{S45}
\end{equation}
With this $\mu_{R}^{\ast}$ the dynamical equation thus becomes
\begin{equation}
\begin{aligned}
\frac{dp_{C}}{dt}= \frac{\omega(k-2)(ck-b)}{k-1}p_{C}(1-p_{C}),
\end{aligned}\tag{S46}
\end{equation}
where the initial fraction of cooperators in the population is denoted by $p_{0}=p_{C}(0)$. The solution of this equation is
\begin{equation}
\begin{aligned}
p_{C}=\frac{1}{1+\frac{1-p_{0}}{p_{0}} e^{-\beta_{\textrm{DB}}t}},
\end{aligned}\tag{S47}
\end{equation}
where $\beta_{\textrm{DB}}=\frac{\omega(k-2)(ck-b)}{k-1}$. It also means that the dynamical system needs infinitely long time to reach the full cooperation state from the initial $p_0<1$. To avoid it, we suppose that the terminal state $p_{C}(t_{f})$ is $1-\delta$, where $\delta$ is the parameter determining the cooperation level at the terminal time. Due to $b/c<k$, $p_{C}$ increases monotonically over time $t$, which leads to $p_{C}(t_{f})> p_{0}$.

Furthermore, the cumulative cost required by the optimal rewarding level $\mu_{R}^{\ast}$ for the dynamical system to reach the expected terminal state $p_{C}(t_{f})$ becomes
\begin{equation}
\begin{aligned}
J_{R}^*=\frac{(kN\mu^{\ast}_{R})^{2}}{2\beta_{\textrm{DB}}}[p_{0}-1+\delta+\ln(\frac{1-p_0}{\delta})].
\end{aligned}\tag{S48}
\end{equation}

The optimal control problem for punishment can be solved similarly and for the $\mu_{P}^{\ast}$ optimal level we have
\begin{equation}
\begin{aligned}
\mu_{P}^{\ast}=\frac{2}{k}(ck-b)
\end{aligned}\tag{S49}
\end{equation}
and
\begin{equation}
\begin{aligned}
p_{C}=\frac{1}{1+\frac{1-p_{0}}{p_{0}} e^{-\beta_{\textrm{DB}}t}}.
\end{aligned}\tag{S50}
\end{equation}
Hence, the cumulative cost produced by the optimal punishing protocol $\mu_{P}^{\ast}$ becomes
\begin{equation}
\begin{aligned}
J_{P}^\ast= \frac{(kN\mu_{P}^{\ast})^{2}}{2\beta_{\textrm{DB}}}[p_{0}-1+\delta+\ln(\frac{1-\delta}{p_0})].
\end{aligned}\tag{S51}
\end{equation}

From these results we can conclude that the optimal levels of negative and positive incentives are identical, i.e., $\mu_{R}^{\ast}=\mu_{P}^{\ast}$, but their cumulative costs could be different. For a proper comparison we can calculate their difference, which is
\begin{equation}
\begin{aligned}
J_{R}^\ast-J_{P}^\ast=\frac{(kN\mu_{v}^{\ast})^{2}}{2\beta_{\textrm{DB}}}\ln[\frac{p_0(1-p_0)}{\delta(1-\delta)}]\,,
\end{aligned}\tag{S52}
\end{equation}
where $\mu^{\ast}_{v}=\mu^{\ast}_{R}=\mu^{\ast}_{P}$. In this work we assume that $p_0>0$, since we do not consider behavioral mutations or errors of strategy updating. In addition, $p_0<p_C(t_{f})=1-\delta$ and we have $p_0+\delta<1$. Thus when $\delta<p_0$, we have $J_{R}^*>J_{P}^\ast$, which means that for DB updating the optimal punishment always requires lower cumulative cost than the usage of optimal reward. But when $\delta>p_0$, we have $J_{R}^*<J_{P}^\ast$, which means that for DB updating the optimal reward always requires lower cumulative cost than the usage of optimal punishment.  This observations are supported by numerical calculations and Monte Carlo simulations as plotted in figure~4 and figure~S5, respectively.

\vbox{}
\leftline{\textbf{2. \,\, BD Updating}}
\noindent \textbf{2.1. \,\, Positive Incentive}\\
According to BD update rule~\cite{ohtsuki_h06, ohtsuki_1}, we randomly choose a focal individual for reproduction proportional to fitness who has $k_{C}$ cooperators and $k_{D}$ defectors among its $k$ neighbors.
If the focal individual adopts strategy $C$, then the fitness of the focal individual is given by
\begin{equation}
\begin{aligned}
f_{C}=1-\omega+\omega[(b-c)k_{C}-c k_{D}].
\end{aligned}\tag{S53}
\end{equation}
Since the offspring of the selected individual replaces one of its neighbors randomly, the probability that $p_{C}$ increases by $1/N$ is
\begin{equation}
\begin{aligned}
P(\Delta p_{C}=\frac{1}{N})=p_{C}\sum_{k_{C}=0}^{k}\binom{k}{k_{C}}(q_{C|C})^{k_{C}}(q_{D|C})^{k_{D}} \frac{f_{C}}{\bar{f}}\frac{k_{D}}{k},
\end{aligned}\tag{S54}
\end{equation}
where $\bar{f}$ represents the average fitness of the whole population. In this case, the number of $CC$--pairs increases by $(k-1)q_{C|D}+1$ and  $p_{CC}$ increases by $[(k-1)q_{C|D}+1]/(kN/2)$ with probability
\begin{equation}
\begin{aligned}
P(\Delta p_{CC}=\frac{(k-1)q_{C|D}+1}{kN/2})=p_{C}\sum_{k_{C}=0}^{k}\binom{k}{k_{C}}(q_{C|C})^{k_{C}}(q_{D|C})^{k_{D}} \frac{f_{C}}{\bar{f}}\frac{k_{D}}{k}.
\end{aligned}\tag{S55}
\end{equation}

In the alternative case, the randomly selected focal individual adopts strategy $D$. Here the fitness of the focal individual is given by
$$
f_{D}=1-\omega+\omega(bk_{C}+0 \cdot k_{D}),
$$ and therefore $p_{C}$ decreases by $1/N$ with probability
\begin{equation}
\begin{aligned}
P(\Delta p_{C}=-\frac{1}{N})=p_{D}\sum_{k_{C}=0}^{k}\binom{k}{k_{C}}(q_{C|D})^{k_{C}}(q_{D|D})^{k_{D}} \frac{f_{D}}{\bar{f}}\frac{k_{C}}{k}.
\end{aligned}\tag{S56}
\end{equation}
Consequently, the number of $CC$--pairs decreases by $(k-1)q_{C|C}$ and $p_{CC}$ decreases by $(k-1)q_{C|C}/(kN/2)$ with probability
\begin{equation}
\begin{aligned}
P(\Delta p_{CC}=-\frac{(k-1)q_{C|C}}{kN/2})=p_{D}\sum_{k_{C}=0}^{k}\binom{k}{k_{C}}(q_{C|D})^{k_{C}}(q_{D|D})^{k_{D}} \frac{f_{D}}{\bar{f}}\frac{k_{C}}{k}.
\end{aligned}\tag{S57}
\end{equation}
Here, the average fitness of the whole population is thus denoted by
\begin{equation}
\begin{aligned}
\bar{f}&= p_{C}\sum_{k_{C}=0}^{k}\binom{k}{k_{C}}(q_{C|C})^{k_{C}}(q_{D|C})^{k_{D}}f_{C}+p_{D}\sum_{k_{C}=0}^{k}\binom{k}{k_{C}}(q_{C|D})^{k_{C}}(q_{D|D})^{k_{D}} f_{D}\\
&=1-\omega+k\omega[(b-c+\mu_{R})p_{CC}+(\mu_{R}-c)p_{CD}+bp_{CD}].
\end{aligned}\tag{S58}
\end{equation}
Based on these calculations, we respectively obtain the time derivatives of $p_{C}$ and $p_{CC}$ as
\begin{equation}
\begin{aligned}
\frac{dp_{C}}{dt}&=\frac{E(\Delta p_{C})}{\Delta t}=\frac{\frac{1}{N}P(\Delta p_{C}=\frac{1}{N})-\frac{1}{N}P(\Delta p_{C}=-\frac{1}{N})}{\frac{1}{N}} \nonumber\\
&=\frac{\omega p_{CD}}{\bar{f}}\{(\mu_{R}-c-b)+(k-1)[(b-c+\mu_{R})q_{C|C}+(\mu_{R}-c)q_{D|C}-b q_{C|D}]\}+o(\omega^{ 2})
\end{aligned}\tag{S59}
\end{equation}
and
\begin{equation}
\begin{aligned}
\frac{dp_{CC}}{dt}&=\frac{E(\Delta p_{CC})}{\Delta t}= \frac{\frac{(k-1)q_{C|D}+1}{kN/2} P(\Delta p_{CC}=\frac{(k-1)q_{C|D}+1}{kN/2})
 -\frac{(k-1)q_{C|C}}{kN/2} P(\Delta p_{CC}=-\frac{(k-1)q_{C|C}}{kN/2})}{\frac{1}{N}} \nonumber\\
&=\frac{2p_{CD}}{k}[(k-1)(q_{C|D}-q_{C|C})+1]+o(\omega).
\end{aligned}\tag{S60}
\end{equation}
Furthermore, we have
\begin{equation}
\begin{aligned}
\frac{dq_{C|C}}{dt}=\frac{2p_{CD}}{kp_{C}}[(k-1)(q_{C|D}-q_{C|C})+1]+o(\omega).
\end{aligned}\tag{S61}
\end{equation}
Hence, the dynamical equation is described by
\begin{equation}
\begin{aligned}
 \left\{\begin{array}{lc}
\frac{dp_{C}}{dt}=\omega \Psi_{\textrm{BD}}^{R}(p_{C},q_{C\mid C})+o(\omega^{ 2}),\nonumber\\
\frac{dq_{C|C}}{dt}=\Phi_{\textrm{BD}}^{R}(p_{C},q_{C\mid C})+o(\omega),
 \end{array}\right.
 \end{aligned}\tag{S62}
\end{equation}
where
\begin{equation*}
\begin{aligned}
 \left\{\begin{array}{lc}
\Psi_{\textrm{BD}}^{R}(p_{C}, q_{C\mid C})=\frac{p_{CD}}{\bar{f}}\{(\mu_{R}-c-b)+(k-1)[(b-c+\mu_{R})q_{C|C}+(\mu_{R}-c)q_{D|C}-b q_{C|D}]\}, \\
\Phi_{\textrm{BD}}^{R}(p_{C}, q_{C\mid C})=\frac{2p_{CD}}{kp_{C}}[(k-1)(q_{C|D}-q_{C|C})+1].
 \end{array}\right.
 \end{aligned}
\end{equation*}
Under weak selection, the velocity of $q_{C|C}$ can be large, and it may rapidly converge to the root defined by $\Phi_{\textrm{BD}}^{R}(p_{C}, q_{C\mid C})=0$ as time $t\rightarrow +\infty$.
Thus, we get
\begin{equation}
\begin{aligned}
 q_{C|C}=p_{C}+\frac{1}{k-1}(1-p_{C}).
\end{aligned}\tag{S63}
\end{equation}
Accordingly, the dynamical equation described by Eq.~(S62) becomes
\begin{equation}
\begin{aligned}
\frac{dp_{C}}{dt}=\frac{\omega k(k-2)(\mu_{R}-c)}{k-1}p_{C}(1-p_{C})+o(\omega^{2}),
\end{aligned}\tag{S64}
\end{equation}
which has two fixed points $p_{C}=0$ and $p_{C}=1$. We define the function $F_{\textrm{BD}}(p_{C}, \mu_{R}, t)$ as
\begin{equation}
\begin{aligned}
F_{\textrm{BD}}(p_{C}, \mu_{R}, t)=\frac{\omega k(k-2)(\mu_{R}-c)}{k-1}p_{C}(1-p_{C})+o(\omega^{2}),
\end{aligned}\tag{S65}
\end{equation}
and the derivative of $F_{\textrm{BD}}(p_{C}, \mu_{R}, t)$ with respect to $p_{C}$ is
\begin{equation}
\begin{aligned}
\frac{dF_{\textrm{BD}}}{dp_{C}}=\frac{\omega k(k-2)(\mu_{R}-c)}{k-1}(1-2p_{C})+o(\omega^{2}).
\end{aligned}\tag{S66}
\end{equation}
For $\mu_{R}>c$, we have $\frac{dF_{\textrm{BD}}}{dp_{C}}|_{{p}_{C}=1}=-\frac{\omega k(k-2)(\mu_{R}-c)}{k-1}<0$ and $\frac{dF_{\textrm{BD}}}{dp_{C}}|_{{p}_{C}=0}=\frac{\omega k(k-2)(\mu_{R}-c)}{k-1}>0$. This implies that the fixed point $p_{C}=1$ is stable and $p_{C}=0$ unstable, i.e., cooperators prevail over defectors. Particularly, when $\mu_{R}=0$, we can see that the fixed point $p_{C}=0$ is always stable and $p_{C}=1$ unstable, which means that cooperation cannot emerge under BD update rule as obtained in Refs.~\cite{ohtsuki_h06, ohtsuki_1}.

\vbox{}
\noindent \textbf{2.2. \,\, Negative Incentive}\\
In this subsection, we consider the negative incentive into the networked Prisoner's Dilemma game with BD updating. According to BD updating, a focal individual is randomly selected for reproduction who has $k_{C}$ cooperators and $k_{D}$ defectors among its $k$ neighbors. Here, we first consider the focal individual adopts strategy $C$.  Then, the fitness of the focal individual is given by
\begin{equation}
\begin{aligned}
f_{C}=1-\omega+\omega[k_{C}(b-c)+k_{D}(-c)],
\end{aligned}\tag{S67}
\end{equation}
 and therefore $p_{C}$ increases by $1/N$ with probability
\begin{equation}
\begin{aligned}
P(\Delta p_{C}=\frac{1}{N})=p_{C}\sum_{k_{C}=0}^{k}\binom{k}{k_{C}}(q_{C|C})^{k_{C}}(q_{D|C})^{k_{D}} \frac{f_{C}}{\bar{f}}\frac{k_{D}}{k},
\end{aligned}\tag{S68}
\end{equation}
where $\bar{f}$ denotes the average fitness of the whole population.
And the number of $CC$--pairs increases by $(k-1)q_{C|D}+1$ and therefore $p_{CC}$ increases by $[(k-1)q_{C|D}+1]/(kN/2)$ with probability
\begin{equation}
\begin{aligned}
P(\Delta p_{CC}=\frac{(k-1)q_{C|D}+1}{kN/2})=p_{C}\sum_{k_{C}=0}^{k}\binom{k}{k_{C}}(q_{C|C})^{k_{C}}(q_{D|C})^{k_{D}} \frac{f_{C}}{\bar{f}}\frac{k_{D}}{k}.
\end{aligned}\tag{S69}
\end{equation}

%First, update a $D$-individual who is replaced by a $C$-neighbor. Let a $C$-neighbor be the focal individual who has $k_{C}$ $C$-neighbors and $k_{D}$ $D$-neighbors with $k_{D}+k_{C}=k$. The probability of such a configuration is $\frac{k!}{k_{C}!k_{D}!}(q_{C|C})^{k_{C}}(q_{D|C})^{k_{D}}$.

%Then, update a $C$-individual who is replaced by a $D$-neighbor. Let a $D$-neighbor be the focal individual who has $k_{C}$ $C$-neighbors and $k_{D}$ $D$-neighbors with $k_{D}+k_{C}=k$ and the probability of this configuration is
%$\frac{k!}{k_{C}!k_{D}!}(q_{C|D})^{k_{C}}(q_{D|D})^{k_{D}}$.

In addition, we consider another case, that is, the randomly selected focal individual adopts strategy $D$. In this case, the fitness of the focal individual is given by
$$
f_{D}=1-\omega+\omega[k_{C}(b-\mu_{P})+k_{D}(-\mu_{P})],
$$ and therefore $p_{C}$ decreases by $1/N$ with probability
\begin{equation}
\begin{aligned}
P(\Delta p_{C}=-\frac{1}{N})=p_{D}\sum_{k_{C}=0}^{k}\binom{k}{k_{C}}(q_{C|D})^{k_{C}}(q_{D|D})^{k_{D}} \frac{f_{D}}{\bar{f}}\frac{k_{C}}{k}.
\end{aligned}\tag{S70}
\end{equation}
And the number of $CC$--pairs decreases by $(k-1)q_{C|C}$ and therefore $p_{CC}$ decreases by
$(k-1)q_{C|C}/(kN/2)$ with probability
\begin{equation}
\begin{aligned}
P(\Delta p_{CC}=-\frac{(k-1)q_{C|C}}{kN/2})=p_{D}\sum_{k_{C}=0}^{k}\binom{k}{k_{C}}(q_{C|D})^{k_{C}}(q_{D|D})^{k_{D}} \frac{f_{D}}{\bar{f}}\frac{k_{C}}{k}.
\end{aligned}\tag{S71}
\end{equation}
Here, the average fitness of whole population can be calculated by
\begin{equation}
\begin{aligned}
\bar{f}&=p_{C}\sum_{k_{C}=0}^{k}\binom{k}{k_{C}}(q_{C|C})^{k_{C}}(q_{D|C})^{k_{D}}f_{C}+p_{D}\sum_{k_{C}=0}^{k}\binom{k}{k_{C}}(q_{C|D})^{k_{C}}(q_{D|D})^{k_{D}} f_{D}\\
&=1-\omega+\omega k[(b-c)p_{CC}-cp_{CD}+(b-\mu_{P})p_{CD}-\mu_{P}p_{DD}].
\end{aligned}\tag{S72}
\end{equation}
From these calculations, we respectively obtain the time derivatives of $p_{C}$ and $p_{CC}$ as
\begin{equation}
\begin{aligned}
\frac{dp_{C}}{dt}&=\frac{E(\Delta p_{C})}{\Delta t}=\frac{\frac{1}{N}P(\Delta p_{C}=\frac{1}{N})-\frac{1}{N}P(\Delta p_{C}=-\frac{1}{N})}{\frac{1}{N}} \nonumber\\
&=\frac{\omega p_{CD}}{\bar{f}}\{\mu_{P}-c-b+(k-1)[(b-c)q_{C|C}-cq_{D|C}+(\mu_{P}-b)q_{C|D}+\mu_{P}q_{D|D}]\}+o(\omega^{2}),
\end{aligned}\tag{S73}
\end{equation}
and
\begin{equation}
\begin{aligned}
\frac{dp_{CC}}{dt}&=\frac{E(\Delta p_{CC})}{\Delta t}= \frac{\frac{(k-1)q_{C|D}+1}{kN/2} P(\Delta p_{CC}=\frac{(k-1)q_{C|D}+1}{kN/2})
 -\frac{(k-1)q_{C|C}}{kN/2} P(\Delta p_{CC}=-\frac{(k-1)q_{C|C}}{kN/2})}{\frac{1}{N}} \nonumber\\
&=\frac{2}{k}p_{CD}[(k-1)(q_{C|D}-q_{C|C})+1]+o(\omega).
\end{aligned}\tag{S74}
\end{equation}
Furthermore, we have
\begin{equation}
\begin{aligned}
\frac{dq_{C|C}}{dt}=\frac{2p_{CD}}{k p_{C}}[(k-1)(q_{C|D}-q_{C|C})+1]+o(\omega).
\end{aligned}\tag{S75}
\end{equation}
Hence, the dynamical equation is described by
\begin{equation}
\begin{aligned}
\left\{\begin{array}{lc}
\frac{dp_{C}}{dt}=\omega \Psi_{\textrm{BD}}^{P}(p_{C}, q_{C\mid C})+o(\omega^{2}),\nonumber\\
frac{dq_{C|C}}{dt}= \Phi_{\textrm{BD}}^{P}(p_{C}, q_{C\mid C})+o(\omega),
 \end{array}\right.
 \end{aligned}\tag{S76}
\end{equation}
where
\begin{equation*}
\begin{aligned}
\left\{\begin{array}{lc}
\Psi_{\textrm{BD}}^{P}(p_{C}, q_{C\mid C})=\frac{p_{CD}}{\bar{f}}\{\mu_{P}-c-b+(k-1)[(b-c)q_{C|C}-cq_{D|C}+(\mu_{P}-b)q_{C|D}+\mu_{P}q_{D|D}]\}, \\
\Phi_{\textrm{BD}}^{P}(p_{C}, q_{C\mid C})=\frac{2p_{CD}}{k p_{C}}[(k-1)(q_{C|D}-q_{C|C})+1].
 \end{array}\right.
 \end{aligned}
\end{equation*}
Under weak selection, the velocity of $q_{C|C}$ can be large, and it may rapidly converge to the root defined by $\Phi_{\textrm{BD}}^{P}(p_{C}, q_{C\mid C})=0$ as time $t\rightarrow +\infty$.
Thus, we get
\begin{equation}
\begin{aligned}
  q_{C|C}= p_{C}+\frac{1}{k-1}(1-p_{C}).
\end{aligned}\tag{S77}
\end{equation}
Accordingly, the system described by Eq.~(S76) becomes
\begin{equation}
\begin{aligned}
\frac{dp_{C}}{dt}= \frac{\omega k(k-2)(\mu_{P}-c)}{k-1}p_{C}(1-p_{C})+o(\omega^{2}),
\end{aligned}\tag{S78}
\end{equation}
which has two fixed points, $p_{C}=0$ and $p_{C}=1$. We define the function $F_{\textrm{BD}}(p_{C}, \mu_{P}, t)$ as
\begin{equation}
\begin{aligned}
F_{\textrm{BD}}(p_{C}, \mu_{P}, t)=\frac{\omega k(k-2)(\mu_{P}-c)}{k-1}p_{C}(1-p_{C})+o(\omega^{2}),
\end{aligned}\tag{S79}
\end{equation}
and the derivative of $F_{\textrm{BD}}(p_{C}, \mu_{P}, t)$ with respect to $p_{C}$ is
\begin{equation}
\begin{aligned}
\frac{dF_{\textrm{BD}}}{dp_{C}}=\frac{\omega k(k-2)(\mu_{P}-c)}{k-1}(1-2p_{C})+o(\omega^{2}).
\end{aligned}\tag{S80}
\end{equation}
For $\mu_{P}>c$, we have $\frac{dF_{\textrm{BD}}}{dp_{C}}|_{{p}_{C}=1}=-\frac{\omega k(k-2)(\mu_{P}-c)}{k-1}<0$ and $\frac{dF_{\textrm{BD}}}{dp_{C}}|_{{p}_{C}=0}=\frac{\omega k(k-2)(\mu_{P}-c)}{k-1}>0$ which implies the fixed point $p_{C}=1$ is stable and $p_{C}=0$ is unstable, i.e., cooperators prevail over defectors. Particularly, when $\mu_{P}=0$, we can see that the fixed point $p_{C}=0$ is always stable and $p_{C}=1$ unstable, which means that cooperation can never emerge under BD update rule as obtained in Refs.~\cite{ohtsuki_h06, ohtsuki_1}.

\vbox{}
\noindent \textbf{2.3. \,\, Optimal Incentive Protocols}\\
In subsections 2.1 and  2.2, we have theoretically obtained the dynamical equation with positive or negative incentive by means of the pair approximation approach in the limit of weak selection, which is given by
\begin{equation}
\begin{aligned}
\frac{dp_{C}}{dt}=F_{\textrm{BD}}(p_{C}, \mu_{v}, t)= \frac{\omega k(k-2)(\mu_{v}-c)}{k-1}p_{C}(1-p_{C})+o(\omega^{2})\,,
\end{aligned}\tag{S81}
\end{equation}
having $p_{C}=0$ and $p_{C}=1$ fixed points. If $\mu_{v}> c$, the former is unstable and the latter is stable, indicating that cooperation will be promoted in the long run. Furthermore, to explore the optimal rewarding and punishing protocols, we now employ the approach of HJB equation to solve the formulated optimal control problems for this BD updating.

First we solve the optimal control problem for rewarding. We define the Hamiltonian function $H_{\textrm{BD}}(p_{C}, \mu_{R}, t)$ as
\begin{equation}
\begin{aligned}
H_{\textrm{BD}}(p_{C}, \mu_{R}, t)=\frac{(kNp_{C}\mu_{R})^{2}}{2}+\frac{\partial J_{R}^{*}}{\partial p_{C}}F_{\textrm{BD}}(p_{C}, \mu_{R}, t),
\end{aligned}\tag{S82}
\end{equation}
where $J_{R}^\ast$ is the optimal cost function of $p_{C}$ and $t$ for the optimal rewarding protocol given as
\begin{equation}
\begin{aligned}
J_{R}^\ast=\int^{t_{f}}_{0}\frac{(kNp_{C} \mu_{R})^{2}}{2}dt.
\end{aligned}\tag{S83}
\end{equation}
Solving $\frac{\partial H_{\textrm{BD}}}{\partial \mu_{R}}=0$, we know that the optimal rewarding protocol $\mu_{R}^{\ast}$ should satisfy
\begin{equation}
\begin{aligned}
\mu_{R}^{\ast}=-\frac{\omega (k-2)(1-p_{C})}{k(k-1)N^{2}p_{C}}\frac{\partial J_{R}^{\ast}}{\partial p_{C}}.
\end{aligned}\tag{S84}
\end{equation}
The corresponding HJB equation  \cite{Evans_05, Geering_07, Lenhart_05} for dynamical system with positive incentive can be written as
\begin{equation}
\begin{aligned}
-\frac{\partial J_{R}^\ast}{\partial t}=H_{\textrm{BD}}(p_{C}, \mu_{R}^{\ast}, t).
\end{aligned}\tag{S85}
\end{equation}
Since we assume that the terminal time $t_{f}$ is not fixed, the optimal cost function $J_{R}^{\ast}(p_{C}, t)$ is independent of $t$. Consequently, we have
\begin{equation}
\begin{aligned}
  \frac{\partial J_{R}^\ast}{\partial t}=0.
\end{aligned}\tag{S86}
\end{equation}
We then obtain
\begin{equation}
\begin{aligned}
\frac{\partial J_{R}^{\ast}}{\partial p_{C}} = 0 \;\; {\rm or} \;\;
\frac{\partial J_{R}^{\ast}}{\partial p_{C}}=-\frac{2N^{2}k(k-1)cp_{C}}{\omega (k-2)(1-p_{C})}.
\end{aligned}\tag{S87}
\end{equation}
As $\mu_{R}>0$ and $p_{C}\in(0, 1)$,  we have
\begin{equation}
\begin{aligned}
\frac{\partial J_{R}^{\ast}}{\partial p_{C}}<0.
\end{aligned}\tag{S88}
\end{equation}
Therefore only $\frac{\partial J_{R}^{\ast}}{\partial p_{C}}=-\frac{2N^{2}k(k-1)cp_{C}}{\omega (k-2)(1-p_{C})}$ holds. By substituting this equation into Eq.~(S84), we obtain the optimal rewarding protocol as
\begin{equation}
\begin{aligned}
  \mu_{R}^{\ast}=2c.
\end{aligned}\tag{S89}
\end{equation}
With the optimal rewarding protocol $\mu_{R}^{\ast}$, the dynamical equation thus becomes
\begin{equation}
\begin{aligned}
\frac{dp_{C}}{dt}=\frac{ \omega k(k-2)c}{k-1}p_{C}(1-p_{C}),
\end{aligned}\tag{S90}
\end{equation}
where the initial fraction of cooperators in the population is denoted by $p_{C}(0)=p_{0}$.
Solving the above equation, we have
\begin{equation}
\begin{aligned}
p_{C}=\frac{1}{1+\frac{1-p_{0}}{p_{0}} e^{-\beta_{\textrm{BD}}t}},
\end{aligned}\tag{S91}
\end{equation}
where $\beta_{\textrm{BD}}=\frac{\omega k(k-2)c}{k-1}$. Hence, the cumulative cost produced by the optimal rewarding protocol is given by
\begin{equation}
\begin{aligned}
J_{R}^\ast=\frac{(kN\mu^{\ast}_{R})^{2}}{2\beta_{\textrm{BD}}}[p_{0}-1+\delta+\ln(\frac{1-p_0}{\delta})].
\end{aligned}\tag{S92}
\end{equation}

If we solve the optimal control problem for punishment, we obtain for the optimal protocol of negative incentive
\begin{equation}
\begin{aligned}
  \mu_{P}^{\ast}= 2c,
\end{aligned}\tag{S93}
\end{equation}
and
\begin{equation}
\begin{aligned}
p_{C}=\frac{1}{1+\frac{1-p_{0}}{p_{0}} e^{-\beta_{\textrm{BD}}t}}.
\end{aligned}\tag{S94}
\end{equation}
Accordingly, the cumulative cost required by the optimal punishing protocol is
\begin{equation}
\begin{aligned}
J_{P}^\ast= \frac{(kN\mu_{P}^{\ast})^{2}}{2\beta_{\textrm{BD}}}[p_{0}-1+\delta+\ln(\frac{1-\delta}{p_0})].
\end{aligned}\tag{S95}
\end{equation}
Therefore the cumulative cost difference of optimal rewarding and punishing protocols is
\begin{equation}
\begin{aligned}
J_{R}^\ast-J_{P}^\ast = \frac{(kN\mu^{\ast}_{v})^{2}}{2\beta_{\textrm{BD}}}\ln[\frac{p_0(1-p_0)}{\delta(1-\delta)}],
\end{aligned}\tag{S96}
\end{equation}
where $\mu^{\ast}_{v}=\mu^{\ast}_{R}=\mu^{\ast}_{P}$. Similarly to the analysis of Eq.~(S52) in subsection~1.3, we also find that $J_{R}^\ast>J_{P}^\ast$ when $\delta<p_0$ and $J_{R}^*<J_{P}^\ast$ when $\delta>p_0$. This implies that for BD updating executing the optimal punishing protocol can induce a lower cumulative cost in comparison with the optimal rewarding one when $\delta<p_0$, and this conclusion is reversed when $\delta>p_0$, which has also been confirmed by numerical calculations and Monte Carlo simulations as presented in figure~4 and figure~S5, respectively.

\vbox{}
\leftline{\textbf{3.\,\,  IM Updating}}
\noindent \textbf{3.1. \,\, Positive Incentive}\\
For IM updating~\cite{ohtsuki_h06, ohtsuki_1}, a focal individual is randomly chosen to update its strategy who has $k_{C}$ cooperators and $k_{D}$ defectors among its $k$ neighbors. If the focal individual adopts strategy $D$, then the fitness of a $C$--neighbor is
\begin{equation}
\begin{aligned}
f_{C}=1-\omega+\omega\{(b-c+\mu_{R})(k-1)q_{C\mid C}+(\mu_{R}-c)[(k-1)q_{D\mid C}+1]\},
\end{aligned}\tag{S97}
\end{equation}
and the fitness of a $D$--neighbor is
\begin{equation}
\begin{aligned}
f_{D}=1-\omega+\omega\{ b(k-1)q_{C\mid D}+0 \cdot [(k-1)q_{D\mid D}+1]\}.
\end{aligned}\tag{S98}
\end{equation}
Besides, the fitness of the focal individual is
\begin{equation}
\begin{aligned}
f_{0}=1-\omega+\omega b k_{C}.
\end{aligned}\tag{S99}
\end{equation}
Since the focal individual can keep its own strategy or imitate a neighbor's strategy with probability proportional to the fitness, the probability that the focal individual adopts strategy $C$ is given by
\begin{equation}
\begin{aligned}
\Theta=\frac{k_{C}f_{C}}{k_{C}f_{C}+k_{D}f_{D}+f_{0}}.
\end{aligned}\tag{S100}
\end{equation}
Therefore, $p_{C}$ increases by $1/N$ with probability
\begin{equation}
\begin{aligned}
 P(\Delta p_{C}=\frac{1}{N})=p_{D}\sum_{k_{C}=0}^{k}\binom{k}{k_{C}} (q_{C|D})^{k_{C}}(q_{D|D})^{k_{D}}\Theta.
\end{aligned}\tag{S101}
\end{equation}
Consequently, the number of $CC$--pairs increases by $k_{C}$ and hence $p_{CC}$ increases by $k_{C}/(kN/2)$ with probability
\begin{equation}
\begin{aligned}
 P(\Delta p_{CC}=\frac{2k_{C}}{kN})=p_{D}\binom{k}{k_{C}}(q_{C|D})^{k_{C}}(q_{D|D})^{k_{D}}\Theta.
\end{aligned}\tag{S102}
\end{equation}

In the alternative case, the randomly selected focal individual adopts strategy $C$. Here the fitness of a $C$--neighbor is
\begin{equation}
\begin{aligned}
f_{C}=1-\omega+ \omega\{(b-c+\mu_{R})[(k-1)q_{C|C}+1]+(\mu_{R}-c)(k-1)q_{D|C}\},
\end{aligned}\tag{S103}
\end{equation}
and the fitness of a $D$--neighbor is
\begin{equation}
\begin{aligned}
f_{D} =1-\omega+ \omega[(k-1)q_{C|D}+1]b.
\end{aligned}\tag{S104}
\end{equation}
Besides, the fitness of the focal individual is
\begin{equation}
\begin{aligned}
f_{0}=1-\omega+\omega[(b-c+\mu_{R})k_{C}+(\mu_{R}-c)k_{D}].
\end{aligned}\tag{S105}
\end{equation}
The probability that the focal individual adopts the strategy $D$ is
\begin{equation}
\begin{aligned}
\Upsilon=\frac{k_{D}f_{D}}{k_{C}f_{C}+k_{D}f_{D}+f_{0}}.
\end{aligned}\tag{S106}
\end{equation}
Thus, $p_{C}$ decreases by $1/N$ with probability
\begin{equation}
\begin{aligned}
 P(\Delta p_{C}=-\frac{1}{N})= p_{C}\sum_{k_{C}=0}^{k} \binom{k}{k_{C}} (q_{C|C})^{k_{C}}(q_{D|C})^{k_{D}}\Upsilon.
\end{aligned}\tag{S107}
\end{equation}
Therefore the number of $CC$--pairs decreases by $k_{C}$ and hence $p_{CC}$ decreases by $k_{C}/(kN/2)$ with probability
\begin{equation}
\begin{aligned}
  P(\Delta p_{CC}=-\frac{2k_{C}}{kN})=p_{C}\binom{k}{k_{C}} (q_{C|C})^{k_{C}}(q_{D|C})^{k_{D}}\Upsilon.
\end{aligned}\tag{S108}
\end{equation}
Based on these calculations, the time derivative of
$p_{C}$ is given by
\begin{equation}
\begin{aligned}
\frac{dp_{C}}{dt}&=\frac{E(\Delta p_{C})}{\Delta t}=\frac{\frac{1}{N}P(\Delta p_{C}=\frac{1}{N})-\frac{1}{N}P(\Delta p_{C}=-\frac{1}{N})}{\frac{1}{N}} \nonumber\\
 &=\frac{\omega k p_{CD}}{(k+1)^{2}}[2(\mu_{R}-c-b)+2(k-1)\xi_{1}+(k-1)(\mu_{R}-c)(q_{C\mid C} \nonumber\\
&+q_{D\mid D})+(k+1)^{2}q_{C\mid C}+q_{D\mid D}\xi_{1}]+o(\omega^{2}),
\end{aligned}\tag{S109}
\end{equation}
where $\xi_{1}=(b-c+\mu_{R})q_{C\mid C}+(\mu_{R}-c)q_{D\mid C}-bq_{C\mid D}$.
Accordingly, the time derivative of $p_{CC}$ is given by
\begin{equation}
\begin{aligned}
\frac{dp_{CC}}{dt}&=\frac{E(\Delta p_{CC})}{\Delta t}= \frac{\sum_{k_{C}=0}^k \frac{2k_{C}}{kN}P(\Delta p_{CC}=\frac{2k_{C}}{kN})- \sum_{k_{C}=0}^k \frac{2k_{C}}{kN}P(\Delta p_{CC}=-\frac{2k_{C}}{kN})}{\frac{1}{N}} \nonumber\\
 &=\frac{2p_{CD}}{k+1}[1+(k-1)(q_{C\mid D}-q_{C\mid C})]+o(\omega).
\end{aligned}\tag{S110}
\end{equation}
Furthermore, we have
\begin{equation}
\begin{aligned}
\frac{dq_{C|C}}{dt}=\frac{2p_{CD}}{(k+1)p_{C}}[1+(k-1)(q_{C\mid D}-q_{C\mid C})]+o(\omega).
\end{aligned}\tag{S111}
\end{equation}
Hence, the dynamical equation is described by
\begin{equation}
\begin{aligned}
 \left\{\begin{array}{lc}
\frac{dp_{C}}{dt}=\omega \Psi_{\textrm{IM}}^{R}(p_{C}, q_{C\mid C})+o(\omega^{ 2}),\nonumber\\
\frac{dq_{C|C}}{dt}=\Phi_{\textrm{IM}}^{R}(p_{C}, q_{C\mid C})+o(\omega),
\end{array}\right.
\end{aligned}\tag{S112}
\end{equation}
where
\begin{equation*}
\begin{aligned}
 \left\{\begin{array}{lc}
\Psi_{\textrm{IM}}^{R}(p_{C}, q_{C\mid C})=\frac{ k p_{CD}}{(k+1)^{2}}[2(\mu_{R}-c-b)+2(k-1)\xi_{1}+(k-1)(\mu_{R}-c)(q_{C\mid C} \nonumber\\
 \qquad  \qquad  \qquad +q_{D\mid D})+(k+1)^{2}q_{C\mid C}+q_{D\mid D}\xi_{1}], \\
\Phi_{\textrm{IM}}^{R}(p_{C}, q_{C\mid C})=\frac{2p_{CD}}{(k+1)p_{C}}[1+(k-1)(q_{C\mid D}-q_{C\mid C})].
\end{array}\right.
\end{aligned}
\end{equation*}
Under weak selection, the velocity of $q_{C|C}$ can be large, and it may rapidly converge to the root defined by $\Phi_{\textrm{IM}}^{R}(p_{C}, q_{C\mid C})=0$ as time $t\rightarrow +\infty$.
Thus, we get
\begin{equation}
\begin{aligned}
 q_{C|C}=p_{C}+\frac{1}{k-1}(1-p_{C}).
\end{aligned}\tag{S113}
\end{equation}
Accordingly, the dynamical equation described by Eq.~(S112) becomes
\begin{equation}
\begin{aligned}
\frac{dp_{C}}{dt}=\frac{\omega k^{2}(k-2)[b+(\mu_{R}-c)(k+2)]}{(k+1)^{2}(k-1)}p_{C}(1-p_{C})+o(\omega^{2}),
\end{aligned}\tag{S114}
\end{equation} which has two fixed points $p_{C}=0$ and $p_{C}=1$. We define the function $F_{\textrm{IM}}(p_{C}, \mu_{R}, t)$ as
\begin{equation}
\begin{aligned}
F_{\textrm{IM}}(p_{C}, \mu_{R}, t)=\frac{\omega k^{2}(k-2)[b+(\mu_{R}-c)(k+2)]}{(k+1)^{2}(k-1)}p_{C}(1-p_{C})+o(\omega^{2}),
\end{aligned}\tag{S115}
\end{equation}
and the derivative of $F_{\textrm{IM}}(p_{C}, \mu_{R}, t)$ with respect to $p_{C}$  is
\begin{equation}
\begin{aligned}
\frac{dF_{\textrm{IM}}}{dp_{C}}=\frac{\omega k^{2}(k-2)[b+(\mu_{R}-c)(k+2)]}{(k+1)^{2}(k-1)}(1-2p_{C})+o(\omega^{2}).
\end{aligned}\tag{S116}
\end{equation}
For $\mu_{R}>c-\frac{b}{k+2}$, we have $\frac{dF_{\textrm{IM}}}{dp_{C}}|_{{p}_{C}=1}=-\frac{\omega k^{2}(k-2)[b+(\mu_{R}-c)(k+2)]}{(k+1)^{2}(k-1)}<0$ and $\frac{dF_{\textrm{IM}}}{dp_{C}}|_{{p}_{C}=0}=\frac{\omega k^{2}(k-2)[b+(\mu_{R}-c)(k+2)]}{(k+1)^{2}(k-1)}>0$ which implies that the fixed point $p_{C}=1$ is stable and $p_{C}=0$ is unstable, i.e.,  cooperators prevail over defectors. Particularly, when $\mu_{R}=0$,  we can see that for $b/c>k+2$, the fixed point $p_{C}=1$ is stable and $p_{C}=0$ unstable. Thus, we obtain the condition $b/c>k+2$ for the evolution of cooperation under IM update rule as previously obtained in Refs.~\cite{ohtsuki_h06, ohtsuki_1}.

\vbox{}
\noindent \textbf{3.2. \,\, Negative Incentive}\\
In this subsection, we consider the negative incentive into the networked Prisoner's Dilemma game with IM updating. According to this rule, we randomly choose a focal individual to update its strategy who has $k_{C}$ cooperators and $k_{D}$ defectors among its $k$ neighbors. If the focal individual adopts strategy $D$, then the fitness of a $C$--neighbor is
\begin{equation}
\begin{aligned}
f_{C}=1-\omega+\omega\{(b-c)(k-1)q_{C\mid C}-c[(k-1)q_{D\mid C}+1]\},
\end{aligned}\tag{S117}
\end{equation}
and the fitness of a $D$--neighbor is
\begin{equation}
\begin{aligned}
f_{D}=1-\omega+\omega\{(b-\mu_{P})(k-1)q_{C\mid D}-\mu_{P}[(k-1)q_{D\mid D}+1]\}.
\end{aligned}\tag{S118}
\end{equation}
Besides, the fitness of the focal individual is
\begin{equation}
\begin{aligned}
f_{0}=1-\omega+\omega[(b-\mu_{P}) k_{C}-\mu_{P} k_{D}].
\end{aligned}\tag{S119}
\end{equation}
The probability that the focal individual adopts strategy $C$ is given by the expression $\Theta$ in Eq.~(S100).
Therefore, $p_{C}$ increases by $1/N$ with probability
\begin{equation}
\begin{aligned}
 P(\Delta p_{C}=\frac{1}{N})=p_{D}\sum_{k_{C}=0}^{k} \binom{k}{k_{C}} (q_{C|D})^{k_{C}}(q_{D|D})^{k_{D}}\Theta.
\end{aligned}\tag{S120}
\end{equation}
Furthermore, the number of $CC$--pairs increases by $k_{C}$ and hence $p_{CC}$ increases by $k_{C}/(kN/2)$ with probability
\begin{equation}
\begin{aligned}
 P(\Delta p_{CC}=\frac{2k_{C}}{kN})=p_{D}\binom{k}{k_{C}} (q_{C|D})^{k_{C}}(q_{D|D})^{k_{D}}\Theta.
\end{aligned}\tag{S121}
\end{equation}

In addition, we consider another case, that is, the randomly selected focal individual adopts strategy $C$. In this case, the fitness of a $C$--neighbor is
\begin{equation}
\begin{aligned}
f_{C}=1-\omega+ \omega\{(b-c)[(k-1)q_{C|C}+1]-c(k-1)q_{D|C}\},
\end{aligned}\tag{S122}
\end{equation}
and the fitness of a $D$--neighbor is
\begin{equation}
\begin{aligned}
f_{D}=1-\omega+ \omega\{(b-\mu_{P})[(k-1)q_{C|D}+1]-\mu_{P}(k-1)q_{D|D}\}.
\end{aligned}\tag{S123}
\end{equation}
Besides, the fitness of the focal individual is
\begin{equation}
\begin{aligned}
f_{0}=1-\omega+\omega[(b-c)k_{C}-ck_{D}].
\end{aligned}\tag{S124}
\end{equation}
The probability that the focal individual adopts strategy $D$ is given by the expression $\Upsilon$ in Eq.~(S106).
Therefore, $p_{C}$ decreases by $1/N$ with probability
\begin{equation}
\begin{aligned}
 P(\Delta p_{C}=-\frac{1}{N})= p_{C}\sum_{k_{C}=0}^{k} \binom{k}{k_{C}} (q_{C|C})^{k_{C}}(q_{D|C})^{k_{D}}\Upsilon.
\end{aligned}\tag{S125}
\end{equation}
And the number of $CC$--pairs decreases by $k_{C}$ and hence $p_{CC}$ decreases by $k_{C}/(kN/2)$ with probability
\begin{equation}
\begin{aligned}
  P(\Delta p_{CC}=-\frac{2k_{C}}{kN}) = p_{C}\binom{k}{k_{C}}(q_{C|C})^{k_{C}}(q_{D|C})^{k_{D}}\Upsilon.
\end{aligned}\tag{S126}
\end{equation}
Based on these calculations, we obtain the time derivative of
$p_{C}$ given by
\begin{equation}
\begin{aligned}
\frac{dp_{C}}{dt}&=\frac{E(\Delta p_{C})}{\Delta t}=\frac{\frac{1}{N}P(\Delta p_{C}=\frac{1}{N})-\frac{1}{N}P(\Delta p_{C}=-\frac{1}{N})}{\frac{1}{N}} \nonumber\\
 &= \frac{\omega kp_{CD}}{(k+1)^{2}}\{2[(\mu_{P}-c-b)+(k-1)\xi_{2}]+(k-1)(\mu_{P}-c)(q_{C\mid C}+q_{D\mid D}) \nonumber\\
&+(k+1)^{2}(q_{C\mid C}+q_{D\mid D})\xi_{2}\}+o(\omega^{2}),
\end{aligned}\tag{S127}
\end{equation}
where $\xi_{2}=(b-c)q_{C\mid C}-cq_{D\mid C}+(\mu_{P}-b)q_{C\mid D}+\mu_{P}q_{D\mid D}$.
Accordingly, the time derivative of $p_{CC}$ is given by
\begin{equation}
\begin{aligned}
\frac{dp_{CC}}{dt}&=\frac{E(\Delta p_{CC})}{\Delta t}= \frac{\sum_{k_{C}=0}^k \frac{2k_{C}}{kN}P(\Delta p_{CC}=\frac{2k_{C}}{kN})- \sum_{k_{C}=0}^k \frac{2k_{C}}{kN}P(\Delta p_{CC}=-\frac{2k_{C}}{kN})}{\frac{1}{N}} \nonumber\\
&=\frac{2p_{CD}}{k+1}[1+(k-1)(q_{C\mid D}-q_{C\mid C})]+o(\omega).
\end{aligned}\tag{S128}
\end{equation}
Furthermore, we have
\begin{equation}
\begin{aligned}
\frac{dq_{C|C}}{dt}=\frac{2p_{CD}}{(k+1)p_{C}}[1+(k-1)(q_{C\mid D}-q_{C\mid C})]+o(\omega).
\end{aligned}\tag{S129}
\end{equation}
Hence, the dynamical equation is described by
\begin{equation}
\begin{aligned}
 \left\{\begin{array}{lc}
\frac{dp_{C}}{dt}=\omega \Psi_{\textrm{IM}}^{P}(p_{C}, q_{C\mid C})+o(\omega^{ 2}),\nonumber\\
\frac{dq_{C|C}}{dt}= \Phi_{\textrm{IM}}^{P}(p_{C}, q_{C\mid C})+o(\omega),
 \end{array}\right.
 \end{aligned}\tag{S130}
\end{equation}
where
\begin{equation*}
\begin{aligned}
 \left\{\begin{array}{lc}
\Psi_{\textrm{IM}}^{P}(p_{C}, q_{C\mid C}) =\frac{k p_{CD}}{(k+1)^{2}}\{2[(\mu_{P}-c-b)+(k-1)\xi_{2}]+(k-1)(\mu_{P}-c)(q_{C\mid C}+q_{D\mid D}) \nonumber\\
\qquad  \qquad  \qquad \quad+(k+1)^{2}(q_{C\mid C}+q_{D\mid D})\xi_{2}\}, \\
\Phi_{\textrm{IM}}^{P}(p_{C}, q_{C\mid C})=\frac{2p_{CD}}{(k+1)p_{C}}[1+(k-1)(q_{C\mid D}-q_{C\mid C})].
 \end{array}\right.
 \end{aligned}
\end{equation*}
Under weak selection, the velocity of $q_{C|C}$ can be large, and it may rapidly converge to the root defined by $ \Phi_{\textrm{IM}}^{P}(p_{C}, q_{C\mid C})=0$ as time $t\rightarrow +\infty$.
Thus, we get
\begin{equation}
\begin{aligned}
 q_{C|C}=p_{C}+\frac{1}{k-1}(1-p_{C}).
\end{aligned}\tag{S131}
\end{equation}
Accordingly, the dynamical equation described by Eq.~(S130) thus becomes
\begin{equation}
\begin{aligned}
\frac{dp_{C}}{dt}=\frac{\omega k^{2}(k-2)[b+(\mu_{P}-c)(k+2)]}{(k+1)^{2}(k-1)}p_{C}(1-p_{C})+o(\omega^{2}),
\end{aligned}\tag{S132}
\end{equation} which has two fixed points $p_{C}=0$ and $p_{C}=1$.
We define the function $F_{\textrm{IM}}(p_{C}, \mu_{P}, t)$ as
\begin{equation}
\begin{aligned}
F_{\textrm{IM}}(p_{C}, \mu_{P}, t)=\frac{\omega k^{2}(k-2)[b+(\mu_{P}-c)(k+2)]}{(k+1)^{2}(k-1)}p_{C}(1-p_{C})+o(\omega^{2}),
\end{aligned}\tag{S133}
\end{equation}
and the derivative of $F_{\textrm{IM}}(p_{C}, \mu_{P}, t)$ with respect to $p_{C}$  is
\begin{equation}
\begin{aligned}
\frac{dF_{\textrm{IM}}}{dp_{C}}=\frac{\omega k^{2}(k-2)[b+(\mu_{P}-c)(k+2)]}{(k+1)^{2}(k-1)}(1-2p_{C})+o(\omega^{2}).
\end{aligned}\tag{S134}
\end{equation}
For $\mu_{P}>c-\frac{b}{k+2}$, we have $\frac{dF_{\textrm{IM}}}{dp_{C}}|_{{p}_{C}=1}=-\frac{\omega k^{2}(k-2)[b+(\mu_{P}-c)(k+2)]}{(k+1)^{2}(k-1)}<0$ and $\frac{dF_{\textrm{IM}}}{dp_{C}}|_{{p}_{C}=0}=\frac{\omega k^{2}(k-2)[b+(\mu_{P}-c)(k+2)]}{(k+1)^{2}(k-1)}>0$, which implies that the fixed point $p_{C}=1$ is stable and $p_{C}=0$ unstable, i.e.,  cooperators prevail over defectors. Particularly, when $\mu_{P}=0$,  we can see that for $b/c>k+2$, the fixed point $p_{C}=1$ is stable and $p_{C}=0$ unstable. Thus, we obtain the condition $b/c>k+2$ for the evolution of cooperation under IM update rule as obtained in Refs.~\cite{ohtsuki_h06, ohtsuki_1}.

\vbox{}
\noindent \textbf{3.3. \,\, Optimal Incentive Protocols}\\
In subsections 3.1 and 3.2, we have theoretically obtained the dynamical equation with positive or negative incentive by means of the pair approximation approach in the limit of weak selection, which is given by
\begin{equation}
\begin{aligned}
\frac{dp_{C}}{dt}=F_{\textrm{IM}}(p_{C}, \mu_{v}, t)=\frac{ \omega k^{2}(k-2)[b+(\mu_{v}-c)(k+2)]}{(k+1)^{2}(k-1)}p_{C}(1-p_{C})+o(\omega^{2}).
\end{aligned}\tag{S135}
\end{equation}
This dynamical equation has two equilibria which are $p_{C}=0$ and $p_{C}=1$. If $\mu_{v}>c-\frac{b}{k+2}$, the former is unstable and the latter is stable, which means that cooperation will be promoted in the long run. Furthermore, to identify the optimal rewarding and punishing protocols, we now use the approach of HJB equation.

The Hamiltonian function for the control problem is
\begin{equation}
\begin{aligned}
H_{\textrm{IM}}(p_{C}, \mu_R, t)=\frac{(k N p_{C}\mu_{R})^{2}}{2}+\frac{\partial J_{R}^{\ast}}{\partial p_{C}}F_{\textrm{IM}}(p_{C}, \mu_{R}, t),
\end{aligned}\tag{S136}
\end{equation}
where $J_{R}^\ast$ is the optimal cost function of $p_{C}$ and $t$ for the optimal rewarding protocol given as
\begin{equation}
\begin{aligned}
 J_{R}^\ast=\int^{t_{f}}_{0}\frac{(kNp_{C}\mu_{R}^{\ast})^{2}}{2}dt.
\end{aligned}\tag{S137}
\end{equation}
Solving $\frac{\partial H_{\textrm{IM}}}{\partial \mu_{R}}=0$, we know that the optimal rewarding protocol $\mu_{R}^{\ast}$ should satisfy
\begin{equation}
\begin{aligned}
 \mu_{R}^{\ast}=-\frac{\omega(k-2)(k+2)(1-p_{C})}{N^{2}(k-1)(k+1)^{2}p_{C}} \frac{\partial J_{R}^{\ast}}{\partial p_{C}}.
\end{aligned}\tag{S138}
\end{equation}
The HJB equation can be written as
\begin{equation}
\begin{aligned}
-\frac{\partial J_{R}^\ast}{\partial t}=H_{\textrm{IM}}(p_{C}, \mu_R^{\ast}, t).
\end{aligned}\tag{S139}
\end{equation}
As the terminal time $t_{f}$ is not fixed, the optimal cost function $J_{R}^{\ast}(p_{C}, t)$ is independent of $t$. Consequently, we have
\begin{equation}
\begin{aligned}
  \frac{\partial J_{R}^\ast}{\partial t}=0.
\end{aligned}\tag{S140}
\end{equation}
We then obtain
\begin{equation}
\begin{aligned}
\frac{\partial J_{R}^{\ast}}{\partial p_{C}} = 0 \;\; {\rm or} \;\;
\frac{\partial J_{R}^{\ast}}{\partial p_{C}}=\frac{2N^{2}p_{C}[b-c(k+2)](k+1)^{2}(k-1)}{\omega (k-2)(k+2)^{2}(1-p_{C})}.
\end{aligned}\tag{S141}
\end{equation}
As $\mu_{R}>0$ and $p_{C}\in(0, 1)$, we have
\begin{equation}
\begin{aligned}
 \frac{\partial J_{R}^{\ast}}{\partial p_{C}}<0.
\end{aligned}\tag{S142}
\end{equation}
From Eq.~(S142), we can see that this inequality is obviously satisfied for $b/c\geq k+2$. Instead, we consider the case, i.e., $b/c<k+2$, and hence only $\frac{\partial J_{R}^{\ast}}{\partial p_{C}}=\frac{2N^{2}p_{C}[b-c(k+2)](k+1)^{2}(k-1)}{\omega (k-2)(k+2)^{2}(1-p_{C})}$ holds. By substituting this equation into Eq.~(S138), we obtain the optimal rewarding protocol as
\begin{equation}
\begin{aligned}
\mu_{R}^{\ast}=\frac{2[c(k+2)-b]}{k+2}.
\end{aligned}\tag{S143}
\end{equation}
With the optimal rewarding protocol $\mu_{R}^{\ast}$, the dynamical equation thus becomes
\begin{equation}
\begin{aligned}
\frac{dp_{C}}{dt}=\frac{\omega k^{2}(k-2)[c(k+2)-b]}{(k+1)^{2}(k-1)}p_{C}(1-p_{C}),
\end{aligned}\tag{S144}
\end{equation}
where the initial fraction of cooperators in the population is denoted by $p_{0}=p_{C}(0)$. To solve the above equation, we have
\begin{equation}
\begin{aligned}
p_{C}=\frac{1}{1+\frac{1-p_{0}}{p_{0}} e^{-\beta_{\textrm{IM}}t}}\,,
\end{aligned}\tag{S145}
\end{equation}
where $\beta_{\textrm{IM}}=\frac{\omega k^{2}(k-2)[c(k+2)-b)]}{(k+1)^{2}(k-1)}$.
Hence, the cumulative cost produced by the optimal rewarding protocol is given by
\begin{equation}
\begin{aligned}
J_{R}^\ast=\frac{(kN\mu^{\ast}_{R})^{2}}{2\beta_{\textrm{IM}}}[p_{0}-1+\delta+\ln(\frac{1-p_0}{\delta})].
\end{aligned}\tag{S146}
\end{equation}

Then, we solve the optimal control problem for punishing described by Eq.~(17) in the main text. After calculations, we respectively obtain the optimal punishing protocol $\mu_{P}^{\ast}$ and the corresponding solution of $p_C$ given by
\begin{equation}
\begin{aligned}
\mu_{P}^{\ast}=\frac{2[c(k+2)-b]}{k+2},
\end{aligned}\tag{S147}
\end{equation}
and
\begin{equation}
\begin{aligned}
p_{C}=\frac{1}{1+\frac{1-p_{0}}{p_{0}} e^{-\beta_{\textrm{IM}}t}}.
\end{aligned}\tag{S148}
\end{equation}
Accordingly, the cumulative cost produced by the optimal punishing protocol is given by
\begin{equation}
\begin{aligned}
J_{P}^\ast=\frac{(kN\mu_{P}^{\ast})^{2}}{2\beta_{\textrm{IM}}}[p_{0}-1+\delta+\ln(\frac{1-\delta}{p_0})].
\end{aligned}\tag{S149}
\end{equation}
Consequently, the cumulative cost difference between optimal rewarding and punishing protocols is given by
\begin{equation}
\begin{aligned}
J_{R}^\ast-J_{P}^\ast = \frac{(kN\mu^{\ast}_{v})^{2}}{2\beta_{\textrm{IM}}}\ln[\frac{p_0(1-p_0)}{\delta(1-\delta)}],
\end{aligned}\tag{S150}
\end{equation}
where $\mu^{\ast}_{v}=\mu^{\ast}_{R}=\mu^{\ast}_{P}$. Similarly to Eq.~(S52), we also find that $J_{R}^\ast>J_{P}^\ast$ when $\delta<p_)$, but when $\delta>p_0$ we have $J_R^*<J_P^*$. This implies that for IM updating the execution of the optimal punishing protocol requires lower cumulative cost in comparison with the optimal rewarding one for $\delta<p_0$ and this conclusion is reversed for $\delta>p_0$. These theoretical results can be confirmed by numerical calculations and Monte Carlo simulations as presented in figure~4 and figure~S5, respectively.

\vbox{}
\leftline{\textbf{4. \,\, PC Updating}}
\noindent \textbf{4.1. \,\, Positive Incentive}\\
For PC updating~\cite{ohtsuki_1} we randomly choose a focal individual to update its strategy who has $k_{C}$ cooperators and $k_{D}$ defectors among its $k$ neighbors. If the focal individual adopts strategy $D$, then the fitness of the focal individual is
\begin{equation}
\begin{aligned}
f_{D}=1-\omega+\omega\pi_{0}^{D}=1-\omega+\omega[bk_{C}+0\cdot k_{D}],
\end{aligned}\tag{S151}
\end{equation}
and the fitness of a $C$--neighbor is
\begin{equation}
\begin{aligned}
f_{C}=1-\omega+\omega\pi_{C}^{D}=1-\omega+\omega\{(b-c+\mu_{R})(k-1)q_{C|C}+(\mu_{R}-c)[(k-1)q_{D|C}+1]\}.
\end{aligned}\tag{S152}
\end{equation}
where $\pi_{0}^{D}$ represents the payoff of the focal individual, and $\pi_{C}^{D}$ denotes the payoff of a $C$--neighbor.

Since the focal individual either keeps its current strategy or adopts the strategy of a neighbor with a probability that depends on the payoff difference, i.e., $\pi_{C}^{D}-\pi_{0}^{D}$, the probability that the focal individual adopts the strategy of a $C$--neighbor for $\omega \rightarrow 0$ is
\begin{equation}
\begin{aligned}
\Lambda=\frac{1}{1+e^{-\omega(\pi_{C}^{D}-\pi_{0}^{D})}}=\frac{1}{2}+\omega\frac{\pi_{C}^{D}-\pi_{0}^{D}}{4}.
\end{aligned}\tag{S153}
\end{equation}
Since $f_{C}-f_{D}=\omega(\pi_{C}^{D}-\pi_{0}^{D})$ for weak selection, we further have
\begin{equation}
\begin{aligned}
\Lambda=\frac{1}{1+e^{-\omega(\pi_{C}^{D}-\pi_{0}^{D})}}=\frac{1}{1+e^{-(f_{C}-f_{D})}}=\frac{1}{2}+\frac{f_{C}-f_{D}}{4}.
\end{aligned}\tag{S154}
\end{equation}
Therefore, $p_{C}$ increases by $1/N$ with probability
\begin{equation}
\begin{aligned}
 P(\Delta p_{C}=\frac{1}{N})=p_{D}\sum_{k_{C}=0}^{k} \binom{k}{k_{C}} (q_{C|D})^{k_{C}}(q_{D|D})^{k_{D}} \frac{k_{C}}{k}\Lambda.
\end{aligned}\tag{S155}
\end{equation}
Hence the number of $CC$--pairs increases by $(k-1)q_{C|D}+1$ and $p_{CC}$ increases by $[(k-1)q_{C|D}+1]/(kN/2)$ with probability
\begin{equation}
\begin{aligned}
 P(\Delta p_{CC}=\frac{(k-1)q_{C|D}+1}{kN/2})=p_{D}\sum_{k_{C}=0}^{k} \binom{k}{k_{C}}(q_{C|D})^{k_{C}}(q_{D|D})^{k_{D}} \frac{k_{C}}{k}\Lambda.
\end{aligned}\tag{S156}
\end{equation}

In addition, we consider another case where the randomly selected focal individual adopts strategy $C$. In this case, the fitness of the focal individual is
\begin{equation}
\begin{aligned}
f_{C}=1-\omega+\omega\pi_{0}^{C}=1-\omega+\omega[(b-c+\mu_{R})k_{C}+(\mu_{R}-c)k_{D}],
\end{aligned}\tag{S157}
\end{equation}
and the fitness of a $D$--neighbor is
\begin{equation}
\begin{aligned}
f_{D}=1-\omega+\omega\pi_{D}^{C}=1-\omega+\omega[(k-1)q_{C|D}+1]b,
\end{aligned}\tag{S158}
\end{equation}
where $\pi_{0}^{C}$ represents the payoff of the focal individual, and $\pi_{D}^{C}$ denotes the payoff of a neighbor with strategy $D$.
The probability that the focal individual adopts the strategy of a $D$--neighbor for $ \omega\rightarrow 0$ is
\begin{equation}
\begin{aligned}
\Omega=\frac{1}{1+e^{-\omega(\pi_{D}^{C}-\pi_{0}^{C})}}=\frac{1}{2}+\omega\frac{\pi_{D}^{C}-\pi_{0}^{C}}{4}=\frac{1}{2}+\frac{f_D-f_C}{4}.
\end{aligned}\tag{S159}
\end{equation}
Therefore, $p_{C}$ decreases by $1/N$ with probability
\begin{equation}
\begin{aligned}
 P(\Delta p_{C}=-\frac{1}{N})=p_{C}\sum_{k_{C}=0}^{k} \binom{k}{k_{C}}(q_{C|C})^{k_{C}}(q_{D|C})^{k_{D}} \frac{k_{D}}{k}\Omega.
\end{aligned}\tag{S160}
\end{equation}
Therefore the number of $CC$--pairs decreases by $(k-1)q_{C|C}$ and hence $p_{CC}$ increases by $(k-1)q_{C|C}/(kN/2)$ with probability
\begin{equation}
\begin{aligned}
 P(\Delta p_{CC}=-\frac{(k-1)q_{C|C}}{kN/2})=p_{C}\sum_{k_{C}=0}^{k} \binom{k}{k_{C}} (q_{C|C})^{k_{C}}(q_{D|C})^{k_{D}} \frac{k_{D}}{k}\Omega.
\end{aligned}\tag{S161}
\end{equation}
Based on these calculations, we respectively obtain the time derivatives of $p_{C}$ and $p_{CC}$ as
\begin{equation}
\begin{aligned}
\frac{dp_{C}}{dt}&=\frac{E(\Delta p_{C})}{\Delta t}=\frac{\frac{1}{N}P(\Delta p_{C}=\frac{1}{N})-\frac{1}{N}P(\Delta p_{C}=-\frac{1}{N})}{\frac{1}{N}} \nonumber\\
 &=\frac{\omega p_{CD}}{2}\{(\mu_{R}-c-b)+(k-1)[(b-c+\mu_{R})q_{C\mid C}+(\mu_{R}-c)q_{D\mid C}-bq_{C\mid D}]\}+o(\omega^{2}).
\end{aligned}\tag{S162}
\end{equation}
and
\begin{equation}
\begin{aligned}
\frac{dp_{CC}}{dt}&=\frac{E(\Delta p_{CC})}{\Delta t}= \frac{\frac{(k-1)q_{C|D}+1}{kN/2} P(\Delta p_{CC}=\frac{(k-1)q_{C|D}+1}{kN/2})
 -\frac{(k-1)q_{C|C}}{kN/2} P(\Delta p_{CC}=-\frac{(k-1)q_{C|C}}{kN/2})}{\frac{1}{N}} \nonumber\\
 &=\frac{1}{k}p_{CD}[1+(k-1)(q_{C\mid D}-q_{C\mid C})]+o(\omega).
\end{aligned}\tag{S163}
\end{equation}
Furthermore, we have
\begin{equation}
\begin{aligned}
\frac{dq_{C|C}}{dt}=\frac{p_{CD}}{k p_{C}}[1+(k-1)(q_{C\mid D}-q_{C\mid C})]+o(\omega).
\end{aligned}\tag{S164}
\end{equation}
Hence, the dynamical equation is described by
\begin{equation}
\begin{aligned}
\left\{\begin{array}{lc}
\frac{dp_{C}}{dt}=\omega \Psi_{\textrm{PC}}^{R}(p_{C}, q_{C\mid C})+o(\omega^{2}),\nonumber\\
\frac{dq_{C|C}}{dt}= \Phi_{\textrm{PC}}^{R}(p_{C}, q_{C\mid C})+o(\omega),
\end{array}\right.
\end{aligned}\tag{S165}
\end{equation}
where
\begin{equation*}
\begin{aligned}
\left\{\begin{array}{lc}
\Psi_{\textrm{PC}}^{R}(p_{C}, q_{C\mid C})=\frac{p_{CD}}{2}\{(\mu_{R}-c-b)+(k-1)[(b-c+\mu_{R})q_{C\mid C}+(\mu_{R}-c)q_{D\mid C}-bq_{C\mid D}]\}\\
\Phi_{\textrm{PC}}^{R}(p_{C}, q_{C\mid C})=\frac{p_{CD}}{k p_{C}}[1+(k-1)(q_{C\mid D}-q_{C\mid C})].
\end{array}\right.
\end{aligned}
\end{equation*}
Under weak selection, the velocity of $q_{C|C}$ can be large, and it may rapidly converge to the root defined by $ \Phi_{\textrm{PC}}^{R}(p_{C}, q_{C\mid C})=0$ as time $t\rightarrow +\infty$.
Thus, we get
\begin{equation}
\begin{aligned}
q_{C|C}=p_{C}+\frac{1}{k-1}(1-p_{C}).
\end{aligned}\tag{S166}
\end{equation}
Accordingly, the dynamical equation described by Eq.~(S165) becomes
\begin{equation}
\begin{aligned}
\frac{dp_{C}}{dt}=\frac{\omega k(k-2)(\mu_{R}-c)}{2(k-1)}p_{C}(1-p_{C})+o(\omega^{2}),
\end{aligned}\tag{S167}
\end{equation} which has two fixed points $p_{C}=0$ and $p_{C}=1$.
We define the function $F_{\textrm{PC}}(p_{C}, \mu_{R}, t)$ as
\begin{equation}
\begin{aligned}
F_{\textrm{PC}}(p_{C}, \mu_{R}, t)=\frac{\omega k(k-2)(\mu_{R}-c)}{2(k-1)}p_{C}(1-p_{C})+o(\omega^{2}),
\end{aligned}\tag{S168}
\end{equation}
and the derivative of $F_{\textrm{PC}}(p_{C}, \mu_{R}, t)$ with respect to $p_{C}$ is
\begin{equation}
\begin{aligned}
\frac{dF_{\textrm{PC}}}{dp_{C}}=\frac{\omega k(k-2)(\mu_{R}-c)}{2(k-1)}(1-2p_{C})+o(\omega^{2}).
\end{aligned}\tag{S169}
\end{equation}
For $\mu_{R}>c$, we have $\frac{dF_{\textrm{PC}}}{dp_{C}}|_{{p}_{C}=1}=-\frac{\omega k(k-2)(\mu_{R}-c)}{2(k-1)}<0$ and $\frac{dF_{\textrm{PC}}}{dp_{C}}|_{{p}_{C}=0}=\frac{\omega k(k-2)(\mu_{R}-c)}{2(k-1)}>0$ which implies that the fixed point $p_{C}=1$ is stable and $p_{C}=0$ is unstable, i.e., cooperators prevail over defectors. Particularly, when $\mu_{R}=0$, we can see that the fixed point $p_{C}=0$ is always stable and $p_{C}=1$ unstable, which means that cooperation can never emerge as observed in previous work \cite{ohtsuki_1}.

\vbox{}
\noindent \textbf{4.2. \,\, Negative Incentive}\\
In this subsection, we consider how punishment works for PC updating. According to this rule, we randomly select a focal individual to update its strategy who has $k_{C}$ cooperators and $k_{D}$ defectors among its $k$ neighbors. If the focal individual adopts strategy $D$, then the fitness of the focal individual is
\begin{equation}
\begin{aligned}
f_{D}=1-\omega+\omega\pi_{0}^{D}=1-\omega+\omega[(b-\mu_{P})k_{C}-\mu_{P} k_{D}],
\end{aligned}\tag{S170}
\end{equation}
and the fitness of a $C$--neighbor is
\begin{equation}
\begin{aligned}
f_{C}=1-\omega+\omega\pi_{C}^{D}=1-\omega+\omega\{(b-c)(k-1)q_{C|C}-c[(k-1)q_{D|C}+1]\},
\end{aligned}\tag{S171}
\end{equation}
where $\pi_{0}^{D}$ represents the payoff of the focal individual, and $\pi_{C}^{D}$ denotes the payoff of a $C$--neighbor.

The probability that the focal individual adopts the strategy of a $C$-neighbor is given by the expression $\Lambda$ in Eq.~(S154).
Therefore, $p_{C}$ increases by $1/N$ with probability
\begin{equation}
\begin{aligned}
 P(\Delta p_{C}=\frac{1}{N})=p_{D}\sum_{k_{C}=0}^{k}\binom{k}{k_{C}} (q_{C|D})^{k_{C}}(q_{D|D})^{k_{D}} \frac{k_{C}}{k}\Lambda,
\end{aligned}\tag{S172}
\end{equation}
and the number of $CC$--pairs increases by $(k-1)q_{C|D}+1$ and hence $p_{CC}$ increases by $[(k-1)q_{C|D}+1]/(kN/2)$ with probability
\begin{equation}
\begin{aligned}
 P(\Delta p_{CC}=\frac{(k-1)q_{C|D}+1}{kN/2})=p_{D}\sum_{k_{C}=0}^{k} \binom{k}{k_{C}}(q_{C|D})^{k_{C}}(q_{D|D})^{k_{D}} \frac{k_{C}}{k}\Lambda.
\end{aligned}\tag{S173}
\end{equation}

In the alternative case, the randomly selected focal individual adopts strategy $C$. Here the fitness of the focal individual is
\begin{equation}
\begin{aligned}
f_{C}=1-\omega+\omega\pi_{0}^{C}=1-\omega+\omega[(b-c)k_{C}-c k_{D}],
\end{aligned}\tag{S174}
\end{equation}
and the fitness of a $D$--neighbor is
\begin{equation}
\begin{aligned}
f_{D}=1-\omega+\omega\pi_{D}^{C}=1-\omega+\omega\{(b-\mu_{P})[(k-1)q_{C|D}+1]-\mu_{P}(k-1)q_{D|D}\},
\end{aligned}\tag{S175}
\end{equation}
where $\pi_{0}^{C}$ represents the payoff of the focal individual, and $\pi_{D}^{C}$ denotes the payoff of a neighbor with strategy $D$.
The probability that the focal individual adopts the strategy of a $D$--neighbor is given by the expression $\Omega$ in Eq.~(S159).
Therefore, $p_{C}$ decreases by $1/N$ with probability
\begin{equation}
\begin{aligned}
 P(\Delta p_{C}=-\frac{1}{N})=p_{C}\sum_{k_{C}=0}^{k} \binom{k}{k_{C}}(q_{C|C})^{k_{C}}(q_{D|C})^{k_{D}} \frac{k_{D}}{k}\Omega.
\end{aligned}\tag{S176}
\end{equation}
And the number of $CC$--pairs decreases by $(k-1)q_{C|C}$ and hence $p_{CC}$ increases by $(k-1)q_{C|C}/(kN/2)$ with probability
\begin{equation}
\begin{aligned}
 P(\Delta p_{CC}=-\frac{(k-1)q_{C|C}}{kN/2})=p_{C}\sum_{k_{C}=0}^{k} \binom{k}{k_{C}}(q_{C|C})^{k_{C}}(q_{D|C})^{k_{D}} \frac{k_{D}}{k}\Omega.
\end{aligned}\tag{S177}
\end{equation}
Based on these calculations, we respectively obtain the time derivatives of $p_{C}$ and $p_{CC}$ as
\begin{equation}
\begin{aligned}
\frac{dp_{C}}{dt}&=\frac{E(\Delta p_{C})}{\Delta t}=\frac{\frac{1}{N}P(\Delta p_{C}=\frac{1}{N})-\frac{1}{N}P(\Delta p_{C}=-\frac{1}{N})}{\frac{1}{N}} \nonumber\\
 &=\frac{\omega p_{CD}}{2}\{\mu_{P}-c-b+(k-1)[(b-c)q_{C\mid C}-cq_{D\mid C}+(\mu_{P}-b)q_{C\mid D}+\mu_{P}q_{D|D}]\}+o(\omega^{2}).
\end{aligned}\tag{S178}
\end{equation}
and
\begin{equation}
\begin{aligned}
\frac{dp_{CC}}{dt}&=\frac{E(\Delta p_{CC})}{\Delta t}= \frac{\frac{(k-1)q_{C|D}+1}{kN/2} P(\Delta p_{CC}=\frac{(k-1)q_{C|D}+1}{kN/2})
 -\frac{(k-1)q_{C|C}}{kN/2} P(\Delta p_{CC}=-\frac{(k-1)q_{C|C}}{kN/2})}{\frac{1}{N}} \nonumber\\
&=\frac{p_{CD}}{k}[1+(k-1)(q_{C\mid D}-q_{C\mid C})]+o(\omega).
\end{aligned}\tag{S179}
\end{equation}
Furthermore, we have
\begin{equation}
\begin{aligned}
\frac{dq_{C|C}}{dt}=\frac{p_{CD}}{kp_{C}}[1+(k-1)(q_{C\mid D}-q_{C\mid C})]+o(\omega).
\end{aligned}\tag{S180}
\end{equation}
Hence, the dynamical equation is described by
\begin{equation}
\begin{aligned}
 \left\{\begin{array}{lc}
\frac{dp_{C}}{dt}=\omega \Psi_{\textrm{PC}}^{P}(p_{C}, q_{C\mid C})+o(\omega^{ 2}),\nonumber\\
\frac{dq_{C|C}}{dt}=\Phi_{\textrm{PC}}^{P}(p_{C}, q_{C\mid C})+o(\omega),
 \end{array}\right.
\end{aligned}\tag{S181}
\end{equation}
where
\begin{equation*}
\begin{aligned}
 \left\{\begin{array}{lc}
\Psi_{\textrm{PC}}^{P}(p_{C}, q_{C\mid C})=\frac{p_{CD}}{2}\{\mu_{P}-c-b+(k-1)[(b-c)q_{C\mid C}-cq_{D\mid C}+(\mu_{P}-b)q_{C\mid D}+\mu_{P}q_{D|D}]\},\\
\Phi_{\textrm{PC}}^{P}(p_{C}, q_{C\mid C})=\frac{p_{CD}}{kp_{C}}[1+(k-1)(q_{C\mid D}-q_{C\mid C})].
 \end{array}\right.
\end{aligned}
\end{equation*}
Under weak selection, the velocity of $q_{C|C}$ can be large, and it may rapidly converge to the root defined by $\Phi_{\textrm{PC}}^{P}(p_{C}, q_{C\mid C})=0$ as time $t\rightarrow +\infty$.
Thus, we get
\begin{equation}
\begin{aligned}
 q_{C|C}=p_{C}+\frac{1}{k-1}(1-p_{C}).
\end{aligned}\tag{S182}
\end{equation}
Accordingly, the dynamical equation described by Eq.~(S181) thus  becomes
\begin{equation}
\begin{aligned}
\frac{dp_{C}}{dt}=\frac{\omega k(k-2)(\mu_{P}-c)}{2(k-1)}p_{C}(1-p_{C})+o(\omega^{2}),
\end{aligned}\tag{S183}
\end{equation} which has two fixed points $p_{C}=0$ and $p_{C}=1$.
We define the function $F_{\textrm{PC}}(p_{C}, \mu_{P}, t)$ as
\begin{equation}
\begin{aligned}
F_{\textrm{PC}}(p_{C}, \mu_{P}, t)=\frac{\omega k(k-2)(\mu_{P}-c)}{2(k-1)}p_{C}(1-p_{C})+o(\omega^{2}),
\end{aligned}\tag{S184}
\end{equation}
and the derivative of $F_{\textrm{PC}}(p_{C}, \mu_{P}, t)$  with respect to $p_{C}$ is
\begin{equation}
\begin{aligned}
\frac{dF_{\textrm{PC}}}{dp_{C}}=\frac{\omega k(k-2)(\mu_{P}-c)}{2(k-1)}(1-2p_{C})+o(\omega^{2}).
\end{aligned}\tag{S185}
\end{equation}
For $\mu_{P}>c$, we have $\frac{dF_{\textrm{PC}}}{dp_{C}}|_{{p}_{C}=1}=-\frac{\omega k(k-2)(\mu_{P}-c)}{2(k-1)}<0$ and $\frac{dF_{\textrm{PC}}}{dp_{C}}|_{{p}_{C}=0}=\frac{\omega k(k-2)(\mu_{P}-c)}{2(k-1)}>0$ which implies that the fixed point $p_{C}=1$ is stable and $p_{C}=0$ unstable, i.e.,  cooperators prevail over defectors. Particularly, when $\mu_{P}=0$, we can see that the fixed point $p_{C}=0$ is always stable and $p_{C}=1$ unstable, which means that cooperation can never emerge as obtained in Ref.~\cite{ohtsuki_1}.

\vbox{}
\noindent \textbf{4.3. \,\, Optimal Incentive Protocols}\\
By means of the pair approximation approach, in the weak selection limit we have the dynamical equation under PC update rule as
\begin{equation}
\begin{aligned}
\frac{dp_{C}}{dt}= F_{\textrm{PC}}(p_{C}, \mu_{v}, t)=\frac{\omega k(k-2)(\mu_{v}-c)}{2(k-1)}p_{C}(1-p_{C})+o(\omega^{2}).
\end{aligned}\tag{S186}
\end{equation}
This dynamical equation has two equilibria which are $p_{C}=0$ and $p_{C}=1$. If $\mu_{v}>c$,  the former is unstable and the latter is stable, and hence cooperation will be promoted in the long run. Furthermore, to explore the optimal rewarding and punishing protocols, we now use the approach of HJB equation.

In case of reward, we define the Hamiltonian function $H_{\textrm{PC}}(p_{C}, \mu_R, t)$ as
\begin{equation}
\begin{aligned}
H_{\textrm{PC}}(p_{C}, \mu_R, t)=\frac{(k N p_{C}\mu_{R})^{2}}{2}+\frac{\partial J_{R}^{\ast}}{\partial p_{C}}F_{\textrm{PC}}(p_{C}, \mu_{R}, t),
\end{aligned}\tag{S187}
\end{equation}
where $J_{R}^\ast$ is the optimal cost function of $p_{C}$ and $t$ for the optimal rewarding protocol given as
\begin{equation}
\begin{aligned}
 J_{R}^\ast=\int^{t_{f}}_{0}\frac{(kNp_{C}\mu_{R}^*)^{2}}{2}dt.
\end{aligned}\tag{S188}
\end{equation}
Solving $\frac{\partial H_{\textrm{PC}}}{\partial \mu_{R}}=0$, we know that the optimal rewarding protocol $\mu_{R}^{\ast}$ should satisfy
\begin{equation}
\begin{aligned}
\mu_{R}^{\ast} =-\frac{\omega (k-2)(1-p_{C})}{2N^{2}k(k-1)p_{C}}\frac{\partial J_{R}^{\ast}}{\partial p_{C}}.
\end{aligned}\tag{S189}
\end{equation}
The HJB equation can be written as
\begin{equation}
\begin{aligned}
-\frac{\partial J_{R}^\ast}{\partial t}=H_{\textrm{PC}}(p_{C}, \mu_{R}^{\ast}, t).
\end{aligned}\tag{S190}
\end{equation}
As the terminal time $t_{f}$ is not fixed, the optimal cost function $J_{R}^{\ast}(p_{C}, t)$ is independent of $t$. Consequently, we have
\begin{equation}
\begin{aligned}
\frac{\partial J_{R}^\ast}{\partial t}=0.
\end{aligned}\tag{S191}
\end{equation}
We then yield
\begin{equation}
\begin{aligned}
\frac{\partial J_{R}^{\ast}}{\partial p_{C}} = 0 \;\; {\rm or} \;\;
\frac{\partial J_{R}^{\ast}}{\partial p_{C}}=-\frac{4N^{2}k(k-1)cp_{C}}{\omega(k-2)(1-p_{C})}.
\end{aligned}\tag{S192}
\end{equation}
As $\mu_{R}>0$ and $p_{C}\in(0, 1)$, we have
\begin{equation}
\begin{aligned}
 \frac{\partial J_{R}^{\ast}}{\partial p_{C}}<0.
\end{aligned}\tag{S193}
\end{equation}
Therefore only $\frac{\partial J_{R}^{\ast}}{\partial p_{C}}=-\frac{4N^{2}k(k-1)cp_{C}}{\omega(k-2)(1-p_{C})}$ holds. By substituting this equation into Eq.~(S189), we obtain the optimal rewarding protocol $\mu_{R}^{\ast}$ as
\begin{equation}
\begin{aligned}
   \mu_{R}^{\ast}=2c.
\end{aligned}\tag{S194}
\end{equation}
With the optimal rewarding protocol $\mu_{R}^{\ast}$, the dynamical equation thus becomes
\begin{equation}
\begin{aligned}
\frac{dp_{C}}{dt}= \frac{\omega k(k-2)c}{2(k-1)}p_{C}(1-p_{C}),
\end{aligned}\tag{S195}
\end{equation}
where the initial fraction of cooperators in the population is denoted by $p_{0}=p_{C}(0)$.
By solving this equation, we have
\begin{equation}
\begin{aligned}
p_{C}=\frac{1}{1+\frac{1-p_{0}}{p_{0}} e^{-\beta_{\textrm{PC}}t}},
\end{aligned}\tag{S196}
\end{equation}
where $\beta_{\textrm{PC}}=\frac{\omega k(k-2)c}{2(k-1)}$.
Hence, the cumulative cost produced by the optimal rewarding protocol is given by
\begin{equation}
\begin{aligned}
J_{R}^*=\frac{(kN\mu^{\ast}_{R})^{2}}{2\beta_{\textrm{PC}}}[p_{0}-1+\delta+\ln(\frac{1-p_0}{\delta})].
\end{aligned}\tag{S197}
\end{equation}

For punishment, we respectively obtain the optimal protocol $\mu_{P}^{\ast}$ and the corresponding solution of $p_C$ as
\begin{equation}
\begin{aligned}
  \mu_{P}^{\ast}=2c
\end{aligned}\tag{S198}
\end{equation}
and
\begin{equation}
\begin{aligned}
p_{C}=\frac{1}{1+\frac{1-p_{0}}{p_{0}} e^{-\beta_{\textrm{PC}}t}}.
\end{aligned}\tag{S199}
\end{equation}
Accordingly, the cumulative cost produced by the optimal punishing protocol is given by
\begin{equation}
\begin{aligned}
J_{P}^\ast=\frac{(kN\mu_{P}^{\ast})^{2}}{2\beta_{\textrm{PC}}}[p_{0}-1+\delta+\ln(\frac{1-\delta}{p_0})].
\end{aligned}\tag{S200}
\end{equation}
Consequently, the difference between the cumulative cost values is
\begin{equation}
\begin{aligned}
J_{R}^\ast-J_{P}^\ast = \frac{(kN\mu^{\ast}_{v})^{2}}{2\beta_{\textrm{PC}}}\ln[\frac{p_0(1-p_0)}{\delta(1-\delta)}],
\end{aligned}\tag{S201}
\end{equation}
where $\mu^{\ast}_{v}=\mu^{\ast}_{R}=\mu^{\ast}_{P}$. Similarly to Eq.~(S52), we also find that $J_{R}^\ast>J_{P}^\ast$ when $\delta<p_0$, but when $\delta>p_0$ we have $J_R^*<J_P^*$. This implies that for PC updating the usage of optimal punishment requires less cost than the optimal rewarding protocol for $\delta<p_0$ and we have the opposite conclusion for $\delta>p_0$. These theoretical results can be confirmed by numerical calculations and Monte Carlo simulations as presented in figure~4 and figure~S5, respectively.

\vfill \eject

\leftline{\textbf{Supplementary Figures}}

\setcounter{figure}{0}
\begin{figure*}[h]
\begin{center}
\includegraphics[width=16cm]{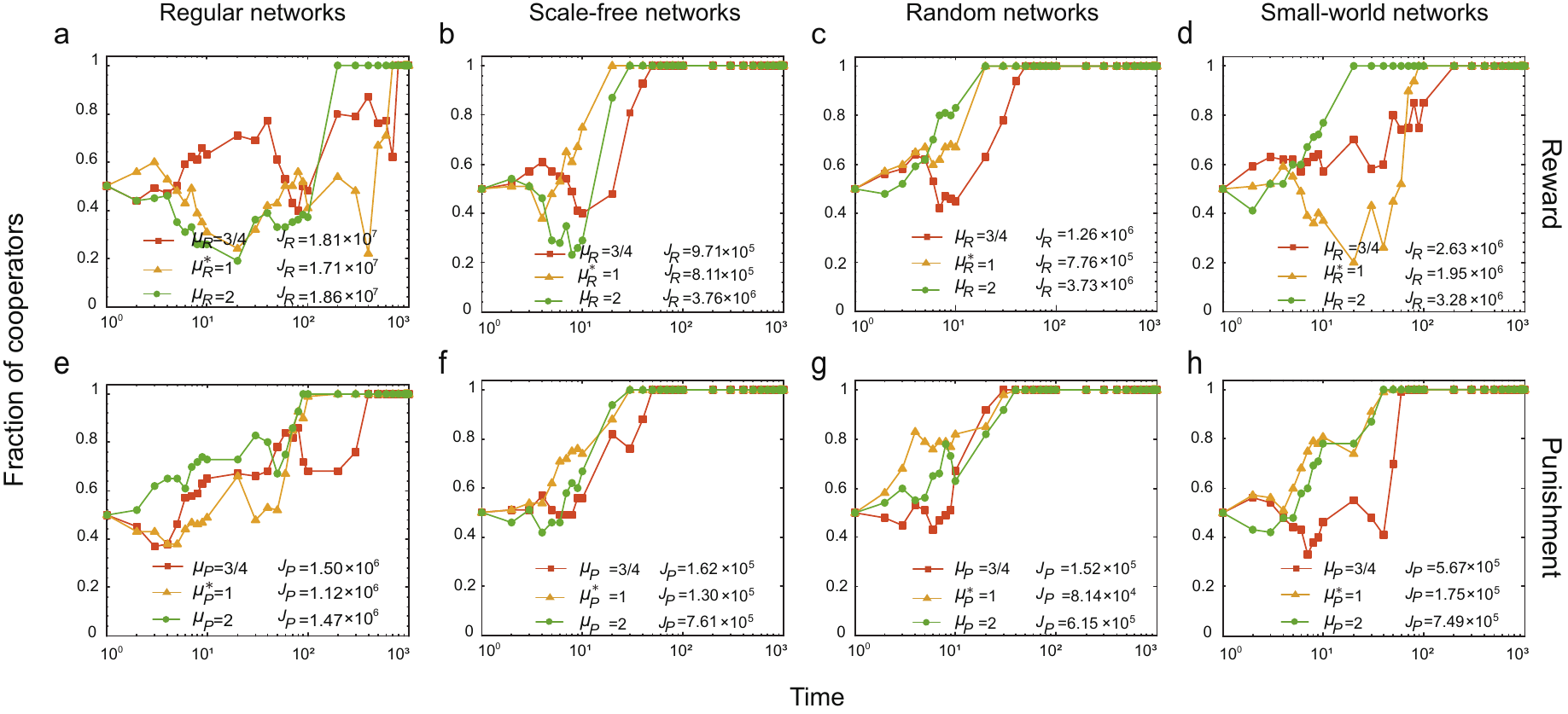}
\caption{\textbf{Time evolution of the fraction of cooperators for three different protocols of incentives on four different networks under DB updating.} The applied protocols are marked by the legend, where the optimal one is indicated by $\ast$. We have also plotted the cumulative cost values for each incentive protocol. The results of Monte Carlo simulations for reward (punishment) are shown on top (bottom) row. Parameters: $N=100$, $L=10$, $b=2$, $c=1$, $\delta=0.01$, $\omega=0.01$, and $p_{0}=0.5$.
For proper comparison the average degree is set to 4 for all graphs.
}\label{figS1}
\end{center}
\end{figure*}
\newpage

\begin{figure*}[!t]
\begin{center}
\includegraphics[width=16cm]{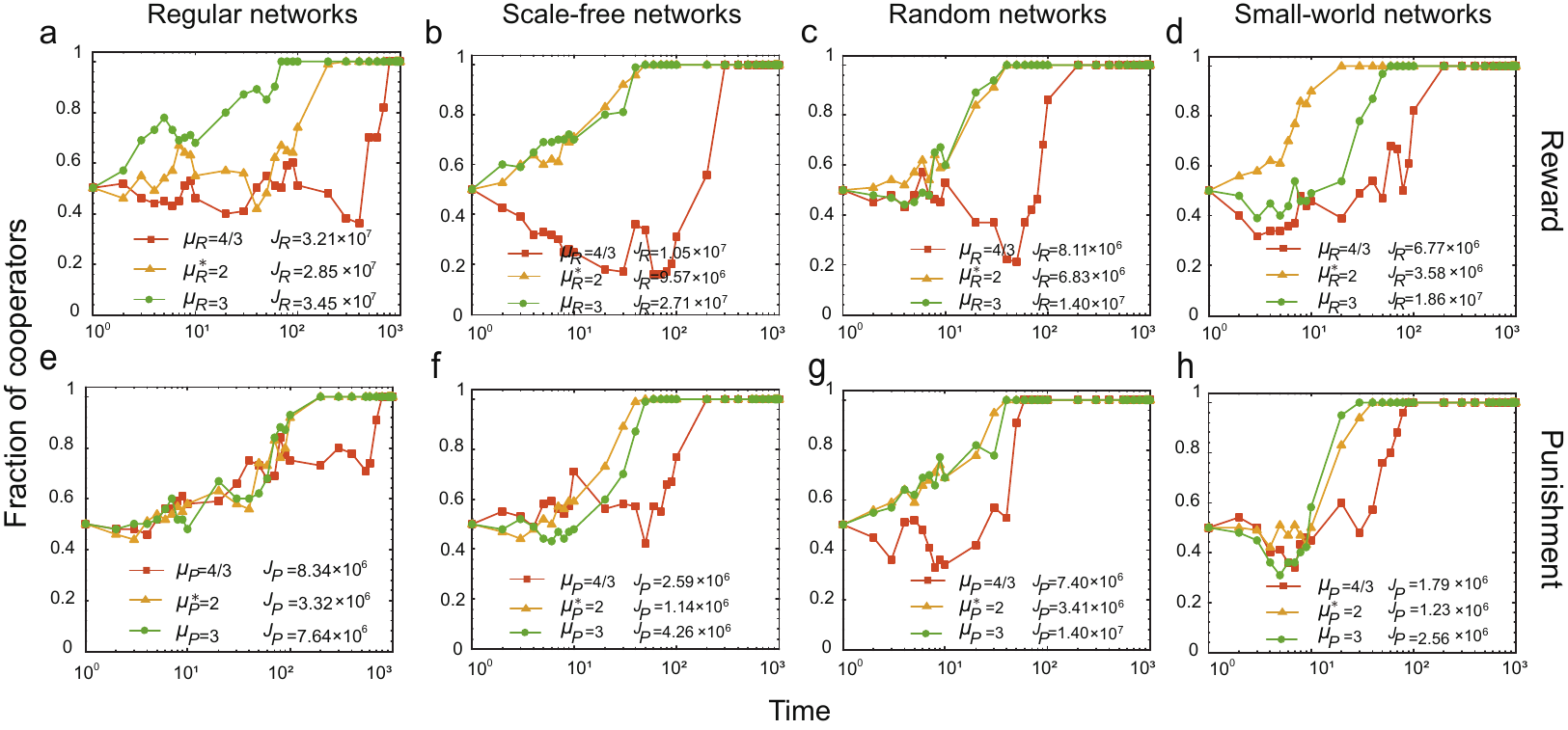}
\caption{\textbf{Time evolution of the fraction of cooperators for three different protocols of incentives on four different networks under BD updating.} The applied protocols are marked by the legend, where the optimal one is indicated by $\ast$. We have also plotted the cumulative cost values for each incentive protocol. The results of Monte Carlo simulations for reward (punishment) are shown on top (bottom) row. Other parameter values are the same as those in figure~S1.}\label{figS2}
\end{center}
\end{figure*}
%\clearpage

\begin{figure*}[!t]
\begin{center}
\includegraphics[width=16cm]{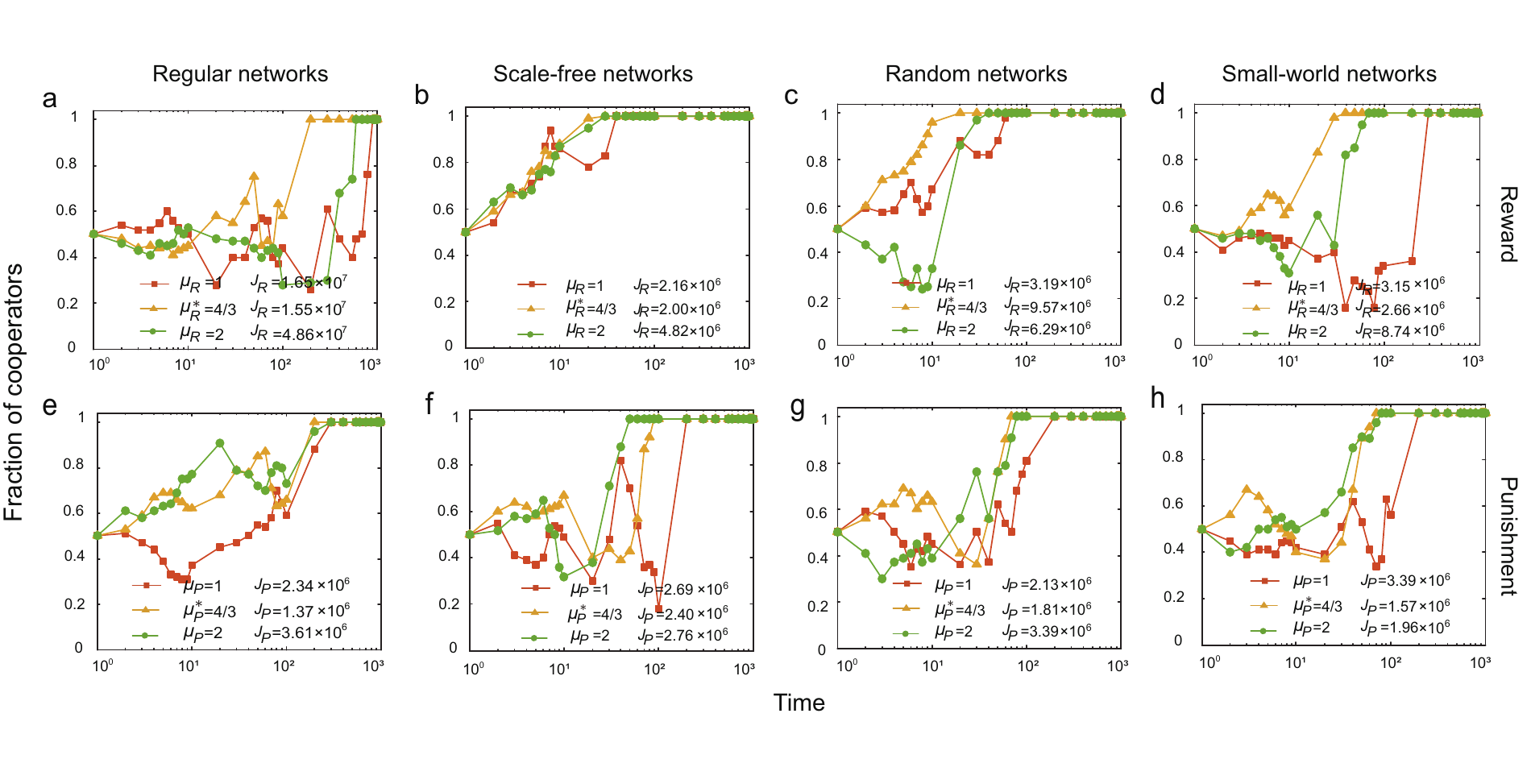}
\caption{\textbf{Time evolution of the fraction of cooperators for three different protocols of incentives on four different networks under IM updating.} The applied protocols are marked by the legend, where the optimal one is indicated by $\ast$. We have also plotted the cumulative cost values for each incentive protocol. The results of Monte Carlo simulations for reward (punishment) are shown on top (bottom) row. Other parameter values are the same as those in figure~S1.}\label{figS3}
\end{center}
\end{figure*}
\clearpage

\begin{figure*}[!t]
\begin{center}
\includegraphics[width=16cm]{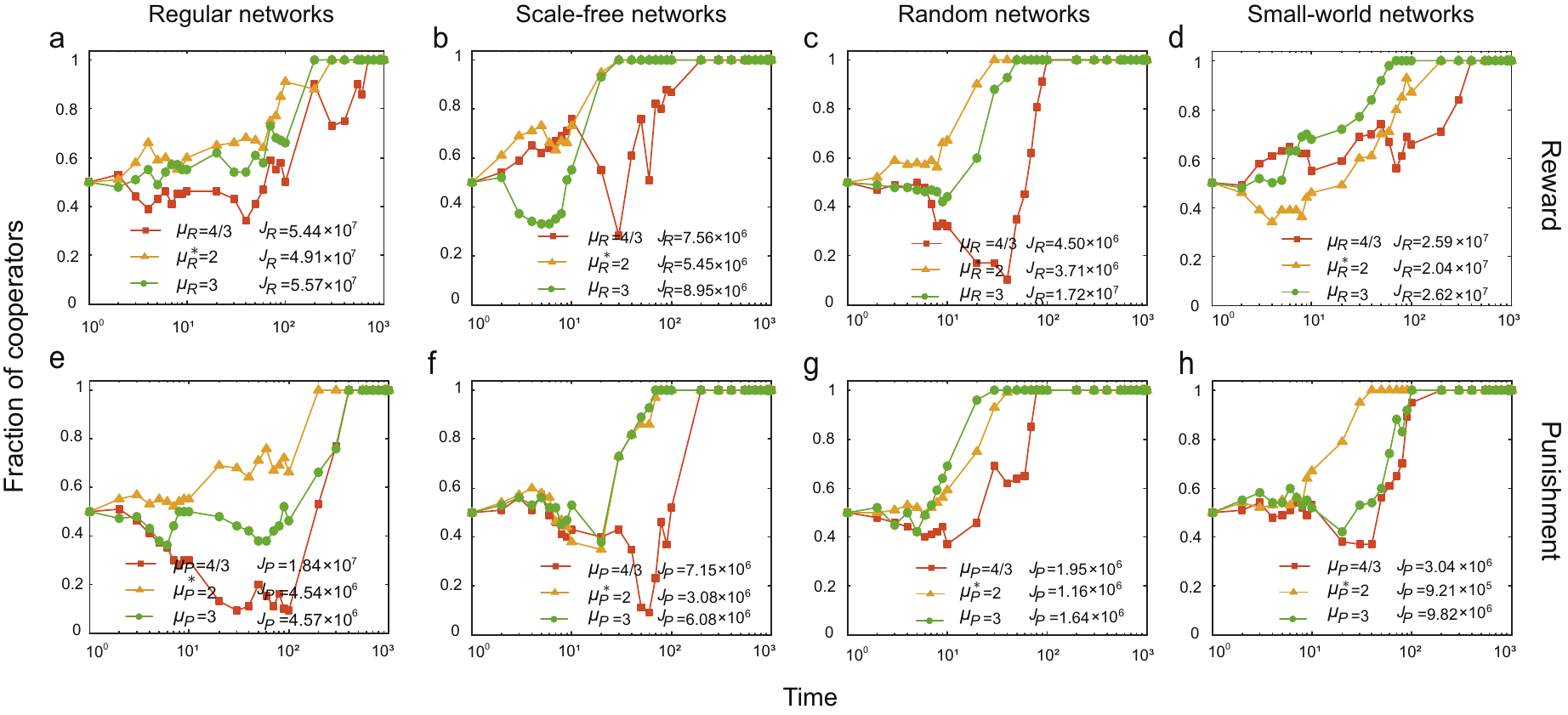}
\caption{\textbf{Time evolution of the fraction of cooperators for three different protocols of incentives on four different networks under PC updating.} The applied protocols are marked by the legend, where the optimal one is indicated by $\ast$. We have also plotted the cumulative cost values for each incentive protocol. The results of Monte Carlo simulations for reward (punishment) are shown on top (bottom) row. Other parameter values are the same as those in figure~S1.}\label{figS4}
\end{center}
\end{figure*}

\begin{figure*}[!t]
\begin{center}
\includegraphics[width=16cm]{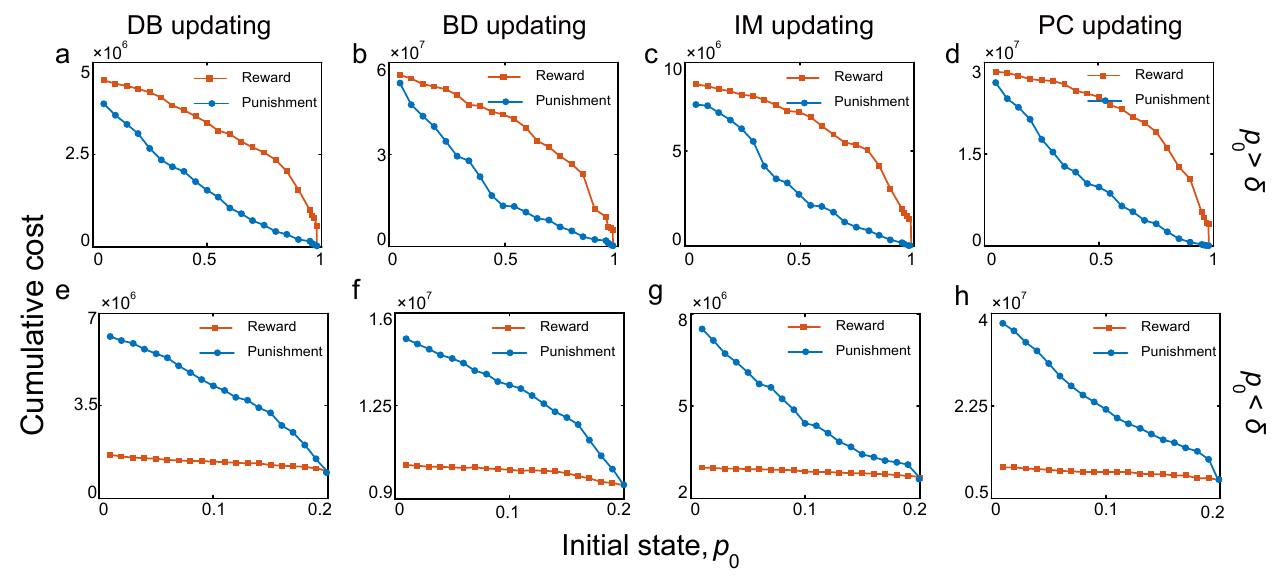}
\caption{\textbf{Cumulative cost needed for reaching the expected terminal state in dependence of the $p_{0}$ initial portion of cooperators for the optimal rewarding and punishing protocols.} Each column of panels represents a strategy update rule as indicated.
Top row represents the results of Monte Carlo simulations by averaging over 200 independent simulation runs on regular networks with degree $k$ in the condition of $p_0>\delta=0.01$, while bottom row represents the results obtained from Monte carlo simulations by averaging over 200 independent simulation runs on regular networks with degree $k$ in the condition of $p_0<\delta=0.2$. Other parameters: $N=100$, $L=10$, $b=2$, $c=1$, $\omega=0.01$, and $k=4$.}\label{figS5}
\end{center}
\end{figure*}

\end{document}